\documentclass[twoside,twocolumn,9pt]{article}
\usepackage{extsizes}
\usepackage[super,sort&compress,comma]{natbib} 
\usepackage[version=3]{mhchem}
\usepackage[left=1.5cm, right=1.5cm, top=1.785cm, bottom=2.0cm]{geometry}
\usepackage{balance}
\usepackage{mathptmx}
\usepackage{sectsty}
\usepackage{graphicx} 
\usepackage{lastpage}
\usepackage[format=plain,justification=justified,singlelinecheck=false,font={stretch=1.125,small,sf},labelfont=bf,labelsep=space]{caption}
\usepackage{float}
\usepackage{fancyhdr}
\usepackage{fnpos}
\usepackage[english]{babel}
\addto{\captionsenglish}{%
  
}
\usepackage{array}
\usepackage{droidsans}
\usepackage{charter}
\usepackage[T1]{fontenc}
\usepackage[usenames,dvipsnames]{xcolor}
\usepackage{setspace}
\usepackage[compact]{titlesec}
\usepackage{hyperref}

\makeFNbottom
\makeatletter
\renewcommand\LARGE{\@setfontsize\LARGE{15pt}{17}}
\renewcommand\Large{\@setfontsize\Large{12pt}{14}}
\renewcommand\large{\@setfontsize\large{10pt}{12}}
\renewcommand\footnotesize{\@setfontsize\footnotesize{7pt}{10}}
\makeatother

\makeatletter 
\renewcommand\@biblabel[1]{#1}            
\renewcommand\@makefntext[1]%
{\noindent\makebox[0pt][r]{\@thefnmark\,}#1}
\makeatother 

\sectionfont{\sffamily\Large}
\subsectionfont{\normalsize}
\subsubsectionfont{\bf}
\titlespacing*{\section}{0pt}{4pt}{4pt}
\titlespacing*{\subsection}{0pt}{15pt}{1pt}

\usepackage{epstopdf}

\definecolor{cream}{RGB}{222,217,201}

\begin{document}

\title{Machine-learning based prediction of small molecule – surface interaction potentials$^\dag$}
\author{Ian Rouse$^{\ast}$\textit{$^{a}$} \and Vladimir Lobaskin \textit{$^{a }$} }
\maketitle
\begin{abstract}
Predicting the adsorption affinity of a small molecule to a target surface is of importance to a range of fields, from catalysis to drug delivery and human safety, but a complex task to perform computationally when taking into account the effects of the surrounding medium. We present a flexible machine-learning approach to predict potentials of mean force (PMFs) and adsorption energies for chemical -- surface pairs from the separate interaction potentials of each partner with a set of probe atoms. We use a pre-existing library of PMFs obtained via atomistic molecular dynamics for a variety of inorganic materials and molecules to train the model. We find good agreement between original and predicted PMFs in both training and validation groups, confirming the predictive power of this approach, and demonstrate the flexibility of the model by producing PMFs for molecules and surfaces outside the training set. 
\end{abstract}

\renewcommand*\rmdefault{bch}\normalfont\upshape
\rmfamily
\section*{}


\footnotetext{\textit{$^{a}$ School of Physics, University College Dublin, Belfield, Dublin 4, Ireland. E-mail: ian.rouse@ucd.ie}}

\footnotetext{\dag Electronic Supplementary Information (ESI) available: Details of material surface and small molecules parameterised for use in the machine learning model, tables of predicted binding energies for sample novel chemicals and surfaces, supplied after main text.}
 
\newcommand{\kjmol}{ $\mathrm{kJ} \cdot \mathrm{mol}^{-1} $}


\section{Introduction}
The interaction between a molecule and an adsorbent surface in a medium is crucial to a wide range of fields of chemistry, ranging from catalysis or drug development to the prediction of toxic effects of nanoparticles (NPs) in the human body \cite{astruc2020introduction, chamundeeswari2019nanocarriers,xia2010index}. Given the high dimensionality of chemical space for both adsorbates and adsorbent surfaces, it is infeasible to experimentally characterise the binding affinity of even a fraction of potential binding partners, while a computational approach based around traditional molecular dynamics simulations would likewise require an impractical amount of time. Docking methods, meanwhile, offer a route to scan large numbers of molecules against target surfaces but are still not strongly developed for molecule--surface rather than molecule-protein systems and in the latter case are known to have significant limitations \cite{bitencourt2019machine,chen2015beware,rodrigues2012metal}. Consequently, there is a need for alternative methods that allow for rapid evaluation of the binding affinity of molecules to surfaces and screening for optimal adsorbate-adsorbent pairs.

On a physical level, the surface and the molecule of interest interact through multiple mechanisms that may include specific or non-specific (electrostatic and van der Waals) forces. In a medium, many-body effects involving solvent molecules might also play a significant role. The solvent mediates van der Waals and electrostatic interactions between the adsorbate and adsorbent and competes with the adsorbate for the place at the surface. Other many-body effects may include hydrophobic attraction between the adsorbent and adsorbate, resulting from the interplay of respective pairwise interactions. All these contributions to the interaction depend on the distance and orientation of the molecule relative to the surface. A full description of the surface-molecule system therefore comprises not only the coordinates of all atoms in the surface and the molecule, but also those of the medium and of the ensemble as a whole.

This complex system can be quantified in terms of a potential of mean force (PMF), which is defined as the free energy profile along a chosen coordinate known as the collective variable and generally obtained via atomistic molecular dynamics using enhanced sampling methods such as metadynamics \cite{kirkwood1935,ensing2006metadynamics,brandt2015molecular,marinova2019time}. The procedure of PMF evaluation involves taking averages over all the remaining degrees of freedom of the medium and the molecule at each value of the collective variable. Thus, the resulting potential includes all the many-body effects and indirect interactions. This operation reduces the large number of degrees of freedom to a more manageable number, typically, a single distance $h$ between the centre of mass (COM) of the adsorbate and the uppermost surface layer of atoms  \cite{brandt2015molecular,marinova2019time} but loses information about the interaction energy at different molecule orientations. The average free energy of adsorption can be obtained from these PMFs according to
\begin{equation} \label{eq:bindingEnergy}
    E_{ads} = - k_B T \ln \left[ \frac{\int_{0}^{\delta_c} \exp \left[-U(h) / k_B T\right] dh }{\int_{0}^{ \delta_c}   dh} \right]
\end{equation}
which parameterises the overall affinity of the molecule to the surface in question by integration of the PMF up to the distance $\delta_c$ at which the molecule is assumed to be unbound.  The PMFs themselves are of use in simulations of more complex systems as part of a multiscale modelling procedure. One particular use is in the prediction of protein-NP adsorption energies in the UnitedAtom model \cite{power2019multiscale}. This requires the set of PMFs for each amino acid side-chain analogue, requiring a significant amount of computational time and producing a set particular to a specific surface geometry and composition. Given the vast array of potential surface-molecule combinations, a more efficient approach for rapidly generating PMFs is of interest. Accurately capturing all the underlying effects in a simple analytical model is not feasible \cite{Horinek2008} and thus we turn to a machine learning (ML) approach for prediction. Many groups have already approached the problem of the binding of ligands to specific targets using ML techniques, as well as the more general cases of the prediction of PMFs, potentials for complex systems or indeed entire forcefields \cite{bertazzo2021machine,gkeka2020machine,dong2010novel,bitencourt2019machine,min2022,schran2021machine,kaser2022neural,Nasikas_2022}, suggesting this is a suitable methodology to apply to the prediction of molecule--surface adsorption.

In principle, it would be possible to simply define a vector of distances $h$ and energies $U(h)$ and either develop a model to predict all of these at once, or to predict these recursively given a known starting point. Most points in the PMF, however, are strongly correlated to those immediately before or after, and so there is a high degree of redundancy if models are developed to predict each point individually. Sequence-based models, e.g. recurrent neural networks,   suffer from the drawback that they typically scale unfavourably with the length of the sequence. The PMFs we are interested in typically cover a range of up to 1.5 nm from the surface at a resolution on the order $0.02$ nm and thus consist of hundreds of datapoints, which would require unfeasible amounts of memory for Transformer based approaches or an exceedingly long runtime for long short-term memory network (LSTM) models \cite{vaswani2017attention}. Thus, we seek a more compact representation of these potentials in order to allow for an efficient predictive model. We must also obtain a suitable set of descriptors to parameterise the surfaces and chemicals to be modelled. The closer these descriptors are to the quantities underlying the molecular dynamics simulations, the more likely we can find a robust model which transfers outside of the training set to novel surfaces and chemicals. To ensure that the methodology is as widely applicable and can be used by as many research groups as possible, these descriptors should not rely on proprietary software and should be able to be calculated in a reasonable timeframe using a typical workstation rather than relying on high-performance computing clusters. Finally, we also require that the descriptors differentiate between different allotropes or crystal phases of the same material and different isomers of the same chemical, and at least potentially be able to describe atoms present in either structure even if they do not appear in the training set. This rules out the use of descriptors which are purely categorical or depend primarily on statistics averaged over atom counts, and suggests that we employ descriptors based on the three-dimensional structure of the compounds involved. Previously, it has been found that interaction potentials offer a useful basis for machine learning of binding affinities \cite{pei2020pair} and we employ a similar approach. Training a model for the prediction of binding energies and, even more so, potentials presents a significant technical challenge as a variety of definitions is used in literature for both the bound state and the distance between the molecule and the surface. Combining data from different sources requires a universal framework to allow a robust mapping of potential profiles and interaction descriptors onto each other. In the following, we describe the proposed procedure in detail.

Here, we present a generic methodology for the prediction of PMFs for small molecules interacting with planar and cylindrical surfaces. Our approach encodes the chemical identity of both the surface and the ligand in terms of their interaction potentials with a set of chemical probes. This representation depends on molecular dynamics forcefield parameters and atomic co-ordinates and as such represents both the component elements and the structure of surfaces and chemicals in a readily extendable manner. These potentials and the target PMFs are converted to a compact set of basis set expansion coefficients using a set of functions based on the hypergeometric functions to minimise the amount of information required to represent them. We employ an artificial neural network implemented using TensorFlow \cite{tensorflow2015-whitepaper}, to convert this representation into the set of coefficients describing the PMF in the same basis set, which provides a smooth analytic function describing the interaction of the small molecule and surface in the medium. The model is trained on results obtained via atomistic molecular dynamics for a range of small organic molecules adsorbing to carbonaceous, metallic, and metal oxide surfaces, with the methodology developed to handle PMFs obtained through multiple computational methods. The predicted PMFs and adsorption energies extracted from these are generally in good agreement with the input values in both training and validation sets. The trained models are incorporated into a suite of Python scripts to form the PMFPredictor Toolkit, which handles the parameterisation of new chemicals and surface structures and the generation of final sets of PMFs, together with scripts to convert chemicals generated using ACPYPE \cite{sousa2012acpype} and surfaces generated using CHARMM-GUI Nanomaterial Modeller \cite{charmmguinanomodeller}.  The entire toolkit including a graphical interface for adding materials is available for download from GitHub \cite{Rouse_PMFPredictor-Toolkit_2022} and the current set of descriptors and predicted PMFs for over $100$ small molecules with over $50$ surfaces archived on Zenodo \cite{rousepmfparchive}. 
 
 \section{Methods}
 \subsection{Overall scheme}
\begin{figure}[tb]
    \centering
    \includegraphics[width=0.48\textwidth]{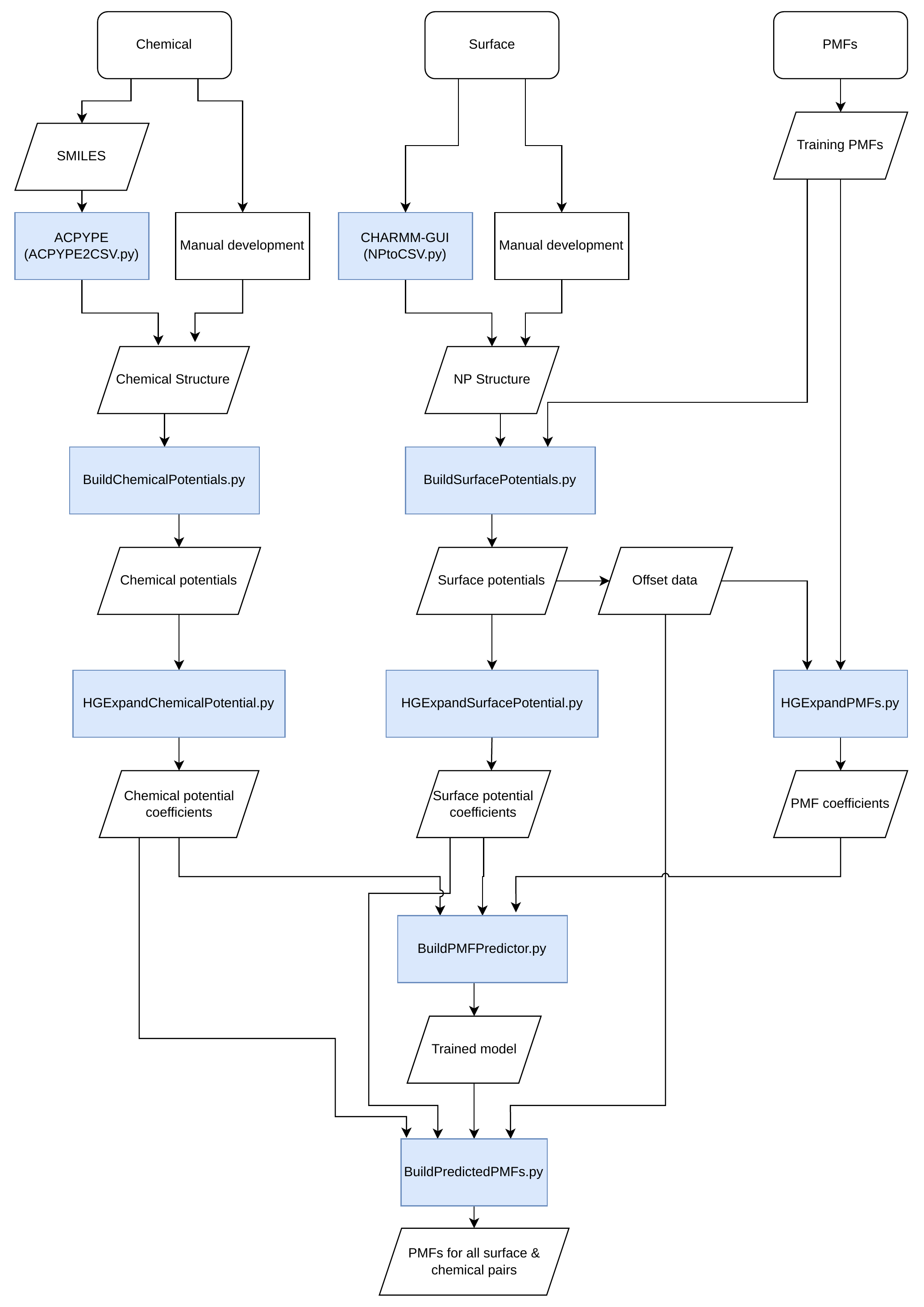}
    \caption{Schematic of the methodology used for the prediction of PMFs. Boxes shaded in blue indicate scripts provided in the toolkit to link different input/output stages together.}
    \label{fig:pmfpredictor-workflow}
\end{figure}

 Briefly, we discuss the overall methodology used for the prediction of PMFs with an overall workflow presented schematically in Fig.~\ref{fig:pmfpredictor-workflow}. We require a flexible means to represent both the material surfaces (hereafter just ``surface'') and molecules of interest (``chemicals'') in a form that can be most easily processed into a PMF.  We have investigated a number of possible descriptors for both, including those obtained via density functional theory, cheminformatics descriptors obtained via the MORDRED server, and machine learning embedding methods \cite{rouse2021,moriwaki2018mordred,guo2016entity}. The most successful has been the description of both surfaces and chemicals in terms of their interaction potentials with a set of probe molecules. Intuitively, it makes sense that these would be quite closely related to the PMF, since this itself is a form of interaction potential, and we demonstrate later that the PMF for a specific molecule--surface pair is not too dissimilar from that of the molecule--surface interaction potential for selected configurations. Moreover, these potentials can be calculated from the structures of the surface or chemical provided a molecular dynamics forcefield is available. For the set of PMFs used to construct the model, this is true for all surfaces and chemicals, and optimised forcefields e.g. INTERFACE \cite{heinz2013thermodynamically} are available to describe a wide range of further surfaces. Even if a highly accurate forcefield is not available, approximate parameters may be used to provide a first estimate of the binding affinity. To reduce the complexity of the input and output, we convert these potentials and the target PMF to a representation in terms of an expansion in terms of a basis set of functions constructed from powers of $1/r$, similar to a classic multipole expansion and enabling the high-resolution potentials to be expressed as a small ($20$) set of expansion coefficients and a characteristic length scale.  This basis set can be expressed in terms of hypergeometric functions and thus we refer to this as the hypergeometric expansion (HGE) method. Finally, a machine-learning based model implemented in TensorFlow is trained to convert the input HGE coefficients describing the interaction potentials for the materials and chemicals into a set of output HGE coefficients describing the PMF. As a training set for this model, we employ sets of PMFs calculated at Stockholm University describing the interactions between small organic molecules -- primarily amino acid side chain analogues and lipid fragments -- parameterised using the GAFF forcefield with face-centered cubic (FCC) gold (100), iron oxide, titanium dioxide in four combinations of crystal phase and surface, silica (amorphous and quartz), cadmium selenide, a variety of functionalised carbon nanotubes (CNTs) and graphene with varying numbers of sheets and surface oxidation, available for download at \cite{lyubartsevbionano} and calculated using methodology as described in \cite{power2019multiscale, rouse2021, saeedimasine2020atomistic}. These are augmented with further PMFs calculated by University College Dublin (UCD) for FCC gold and silver (100), (110) and (111) surfaces , which include additional sugar molecules and use CHARMM parameters to describe the adsorbates \cite{subbotinasilver, subbotinagold,subbotinajpcb}.  Lists of the small molecules and surfaces are provided in the ESI Tables S1 and S4 respectively, see also ESI Figure S1 for structures of the small molecules.  Our methodology is therefore designed to incorporate not only this wide range of structures, which covers cylindrical and planar geometries each with varying degrees of roughness and surface modification, but also different forcefields and conventions for the collective distance variable. The trained model is integrated into a pipeline of scripts which handle the generation of input potentials and output of final PMFs as shown in Fig.~\ref{fig:pmfpredictor-workflow}. Prediction of PMFs for novel chemicals can be achieved using their SMILES code as input to ACPYPE, with a wrapper script provided to convert this output to the format expected for the following script. Alternatively, if a structure and forcefield parameters are already known for this chemical, they can be manually converted to the input format. Novel surfaces likewise require a structure and set of forcefield parameters and we provide a wrapper for converting the output from the CHARMM-GUI Nanomaterial modeller tool to the required input format. We furthermore provide a GUI for convenient access to the set of scripts for generating new input and generating a set of predicted PMFs. Throughout, we use units of nm for distance and \kjmol for energy, with the temperature $T = 300$ K assumed where relevant. We denote point-point distances by $r$, the distance from a point to a reference surface by $d$, and the PMF collective variable by $h$.

\subsection{Material and chemical definition} \label{section:inputDefinition}
During the early development of the model, both the materials and chemicals were defined using a set of descriptors obtained from density functional theory and other methods, augmented with SMILES-based descriptors generated via the MORDRED \cite{moriwaki2018mordred} web interface for chemicals, and further descriptors learnt during model training using an embedding technique. The correlation of the pre-specified descriptors with the adsorption energies was typically quite low, limiting the ability of the model to extrapolate to new materials. Better performance was found when using the embedding technique, but this cannot be applied to chemicals and materials outside the training set. This necessitated the development of descriptors more closely related to the adsorption properties of the chemicals in question. A further challenge is finding a representation of the surface structure of the material, e.g., representing the difference between different Miller indices and crystal structures or realisations of different random surfaces. Likewise, it is necessary to differentiate between different chemicals with the same empirical formula, e.g. the leucine and isoleucine side chain analogues.  To overcome these issues and produce an input which is already similar to the desired output of a potential, we define the surfaces and chemicals in terms of a set of potentials representing their interactions with various probes representing atoms (see Table \ref{tab:AtomPotentialProbes}), a generic planar surface (for chemicals only) and small molecules, taking into account multiple possible orientations of the probe molecule relative to the surface or chemical in question. The atomic probes are selected to characterise three main axes: charge affinity, van der Waals affinity, and length scales by systematically varying the charge, Lennard-Jones (LJ) $\epsilon$ and $\sigma$ parameters, respectively. The small molecules here consist of water, comprised of the O and two HW atoms based on the TIP3P model of water, a rigid model of methane consisting of the C and 4 HC atoms in a tetrahedral configuration based on the structure generated by ACPYPE, a six-membered ring of C atoms, and a line of C atoms consisting of either 3 (for surfaces) or 7 (for molecules) atoms, using the smaller set for surfaces for reasons of computational efficiency and to avoid edge effects. The resulting set of potentials include a representation of the spatial arrangement and chemical identity of all atoms present in the structure (chemical or surface) of interest, and can be calculated for new structures using Python scripts supplied in the repository \cite{Rouse_PMFPredictor-Toolkit_2022}.  

\begin{table}[tb]
    \centering
    \begin{tabular}{cccc}
    \hline
        Label   & $\sigma [\mathrm{nm}]$ & $\epsilon [$\kjmol$]$ &  Charge $[e]$ \\
        \hline
C       & 0.339                          & 0.360                                & 0              \\
K       & 0.314                          & 0.364                                & 1              \\
Cl      & 0.404                          & 0.628                                & -1             \\
C2A     & 0.200                          & 0.360                                & 0              \\
C4A     & 0.400                          & 0.360                                & 0              \\
CPlus   & 0.339                          & 0.360                                & 0.5            \\
CMinus  & 0.339                          & 0.360                                & -0.5           \\
CMoreLJ & 0.339                          & 0.5                                  & 0              \\
CLessLJ & 0.339                          & 0.2                                  & 0             \\
CEps20   & 0.339                          & 20                                  & 0             \\
O*       & 0.339                          & 0.2                                  & 0             \\
HW*      & 0.339                          & 0.2                                  & 0             \\
HC*      & 0.339                          & 0.2                                  & 0             \\
    \end{tabular}
    \caption{A summary of the atomic parameters used to generate descriptive potentials to parameterise molecules and surfaces. Atoms marked with a $*$ are not used as individual probes but included in molecular probes. }
    \label{tab:AtomPotentialProbes}
\end{table}

The potential describing the interaction between a structure and a probe are constructed from two components: the van der Waals potential in the LJ model and the electrostatic potential. We first consider the total LJ potential obtained by summation over all atoms in the structure, indexed $i$, with a point atom defined by parameters $\epsilon_p,\sigma_p$ employing standard mixing rules $\sigma_{ip} = \frac{1}{2} (\sigma_i + \sigma_p)$, $\epsilon_{ip} = \sqrt{\epsilon_i \epsilon_p}$ such that the potential is given by
 \begin{equation}
     U_{LJ,p}(r) = \sum_i 4 \epsilon_{ip} \left[  \left(\frac{\sigma_{ip}}{r_i}\right)^{12} -\left(\frac{\sigma_{ip}}{r_i}\right)^{6} \right],
 \end{equation}
 where $r$ is the location of the point atom relative to the structure as discussed later and $r_i$ is the distance between atom $i$ and the probe atom at $r$. The electrostatic potential is given by the standard form
 \begin{equation}
     U_{el,p}(r) = \sum_i \frac{1}{4\pi \epsilon_r \epsilon_0} \frac{q_i q_p}{ r_{i}} ,
 \end{equation}
where we take $\epsilon_r = 1$, i.e. neglecting the effects of the medium. These contributions are then summed together for each point in the probe, $U_{tot}(r) = \sum_p U_{LJ,p}(r) + U_{el,p}(r)$ and evaluated on a grid of points corresponding to a single value of the reference distance $d$, taking $d$ to be the height above a reference plane (defined later) for a planar structure, the radial distance from a reference surface for a cylindrical structure, or the distance from the COM for a chemical to points in a spherical grid, and where $r$ in the above is a function of $d$. The resolution of the grid used for molecular probes is reduced slightly to compensate for the increased computational time required to evaluate multiple orientations of these. By evaluating the total potential at each point on the grid for a given value of $d$, we extract an effective free energy at this distance by averaging over multiple degrees of freedom according to
\begin{equation} \label{eq:potentialFreeEnergy}
    U_F(d) = - k_B T \mathrm{ln} \left(  \frac{ \int  \mathrm{exp} \left[- U_{tot}(d, \tau)/k_B T \right] d \tau }{ \int d \tau } \right),
\end{equation}
where $\tau$ represents all variables to be averaged over, e.g. those parallel to the surface of a plane, the internal angles defining the orientation of a molecular probe, spherical angles defining the grid surrounding a chemical, etc, and with any necessary weighting functions such as the factor $\sin \theta$ required for averaging over the surface of a sphere implicitly contained in $d \tau$. The limits of  integration for surfaces are chosen to cover a sufficiently large portion of the material surface to capture surface irregularities without approaching the boundaries, since for reasons of speed we do not implement periodic boundary conditions. In certain cases, e.g. charge-charge interactions at short range, the numerical evaluation of Eq. \eqref{eq:potentialFreeEnergy} returns infinite results due to numerical overflow and in these cases we approximate $U_F(d)$ for that probe by the minimum value of the energy at that distance. Since Eq. \eqref{eq:potentialFreeEnergy} is essentially a soft-minimum function, this does not lead to too significant an error and we find the resulting potentials remain smooth despite this approximation. We further record the minimum energy at each value of $d$ for use as a further model input, i.e., Eq. \eqref{eq:potentialFreeEnergy} evaluated in the limit $T \rightarrow 0$. 
 
 For chemicals, we generate an additional potential corresponding to the interaction between the chemical and an infinite slab of number density $\rho_i$,
\begin{equation} \label{eq:ljSlab}
    U_{LJ,c}(d) = \sum_i \rho_i 4 \pi \epsilon_{ip} \sigma_{ip}^3 \left( \frac{2}{45} \left( \frac{\sigma_{ip}}{d_i} \right)^9 - \frac{1}{3} \left ( \frac{\sigma_{ip}}{d_i}\right)^3 \right) .
\end{equation}
where $d_i$ is the minimum distance between atom $i$ and the slab, where the slab is defined by the point $(d \cos \phi \sin \theta, d \sin \phi \sin \theta, d \cos \theta) $ and the normal vector defined by the COM of the molecule to this point, taking the slab to be infinite in the directions perpendicular to this vector and extends infinitely outwards. In general, the number density can be estimated from the proportion of each type of atom in the material, the size of the atom, and the packing fraction $\eta_i$. Here, we assume that the slab consists of a single type of atom with the same LJ parameters as the carbon point probe. The volume per atom is given by $4/3 \pi (\sigma_i/2)^3$ such that $\rho_i = 3/\pi \eta_i /\sigma_i^3$, where $\eta_i \approx 0.74$ as an upper bound. The exact value of this density is not too significant in the present work as the same value is applied for all chemicals, but may play a role if further slab potentials are added.

\begin{figure}[tb]
    \centering
    \includegraphics[width=0.42\textwidth]{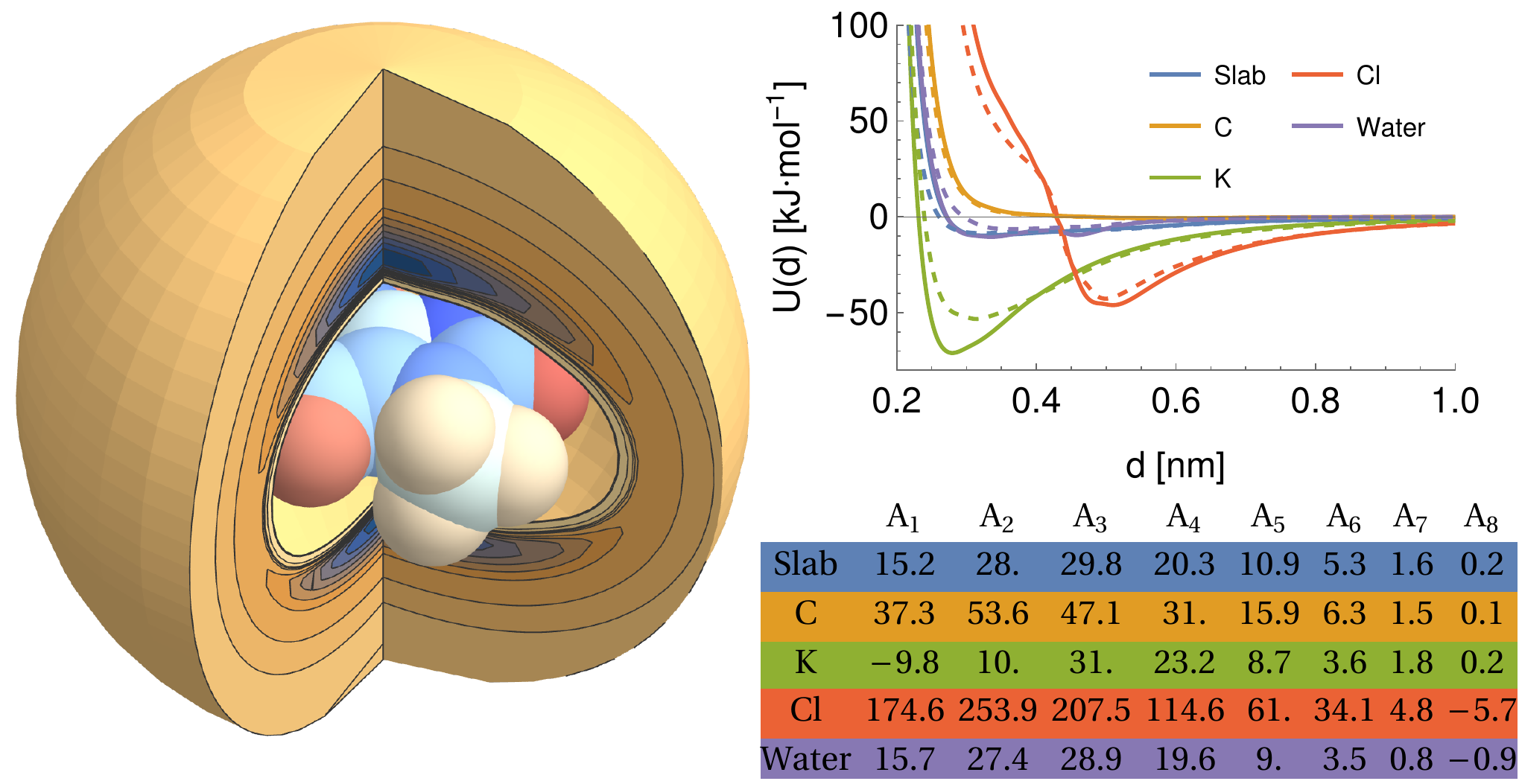}
    \caption{An overview of the procedure for obtaining a numerical representation of a given molecule demonstrated for the tryptophan side-chain analogue (TRPSCA). Left: 3D structure of TRPSCA shown with the potential arising by summing the LJ point-point potential over all atoms present in the molecule with a carbon atom probe. Atoms here are coloured by their partial charge distribution and represented by spheres of radius $\sigma_i/2$. Top right: Potentials generated for a range of probes (see Table \ref{tab:AtomPotentialProbes}) using the GAFF (solid lines) and CHARMM (dashed lines) forcefields to describe TRPSCA. Bottom right: Array of hypergeometric expansion coefficients describing the potentials for the GAFF model of TRPSCA suitable for use in machine-learning models.}
    \label{fig:ljchempointslab}
\end{figure}

To demonstrate this procedure, we show the results obtained for a selection of the atomic and molecular probes  for the tryptophan side-chain analogue (TRPSCA) in Figure \ref{fig:ljchempointslab}, using both CHARMM and GAFF models for the molecule. In this case, we observe that the GAFF model is more strongly interacting overall, which is especially obvious for the interaction with a planar surface and with the potassium ion probe. The primary difference between the two forcefields appears to be the treatment of hydrogen atoms, which in the GAFF model are typically less strongly interacting with both smaller values of $\epsilon$ and $\sigma$ than their equivalents in the CHARMM model, leading to an observable difference in the interaction with neutral atoms. We note also that the charge distribution in the CHARMM model for TRPSCA is more strongly weighted to certain atoms than in GAFF, as exemplified by the nitrogen atom (GAFF $-0.1954 e$ vs CHARMM $-0.5 e$) while maintaining overall neutrality. 

\begin{figure}[tb]
    \centering
    \includegraphics[width=0.42\textwidth]{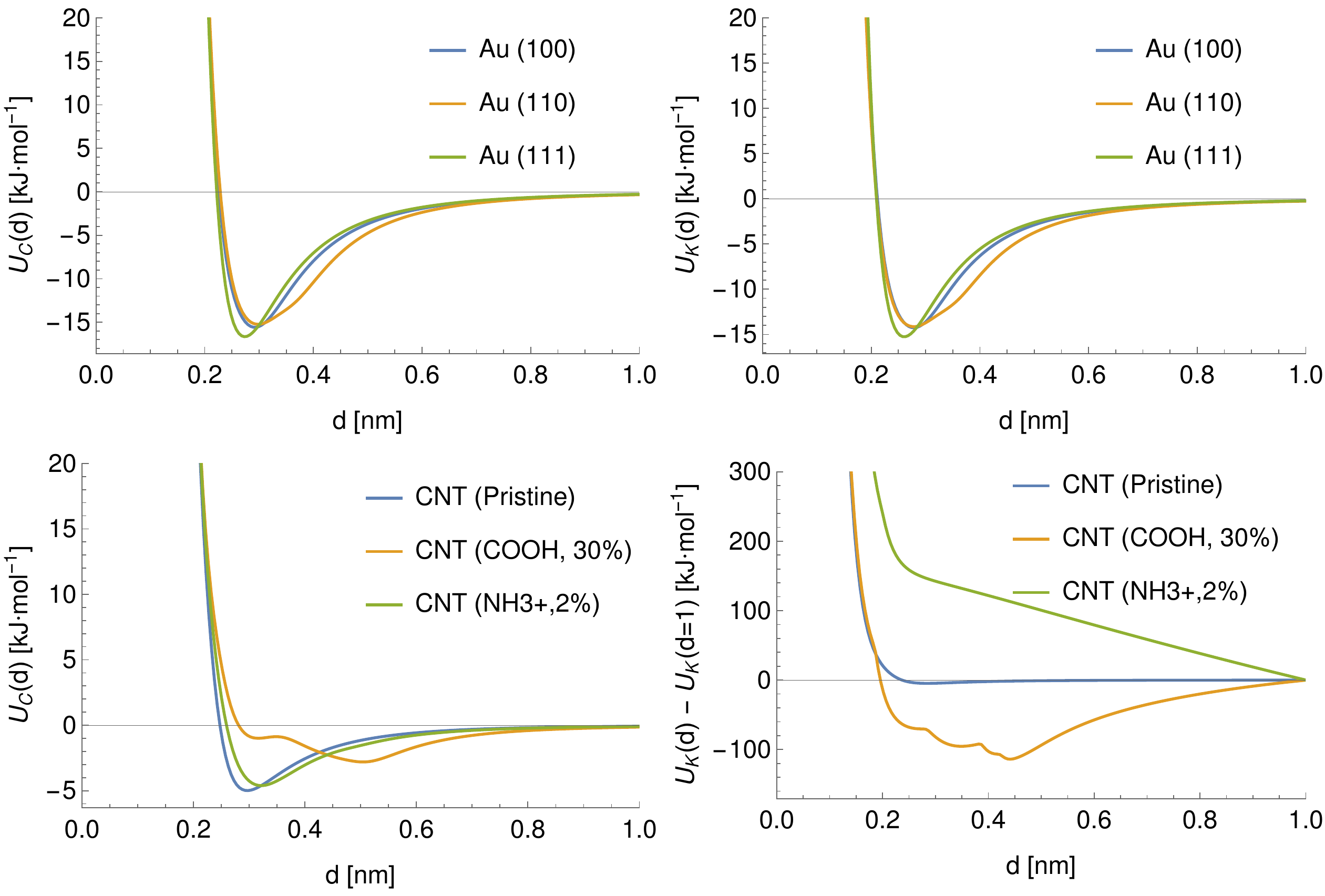}
    \caption{Top row: potentials generated for two point probes, carbon (left) and potassium (right) for three crystal surfaces of FCC gold, showing the (100), (110) and (111) surfaces. Bottom row: As top, except for three types of carbon nanotube: pristine, COOH modified (30 \% by weight) and NH2+ modified (2\% by weight).  }
    \label{fig:matcompare}
\end{figure}

For chemicals, the potential can be defined relative to the COM, which provides a physically meaningful reference point and is straightforward to calculate. The surfaces, however, are infinite along at least one axis and may possess an arbitrary degree of surface roughness or modification. For the cylindrical structures, the distance for all provided PMFs is defined with respect to a fixed radius $R = 0.75$ nm. For the planar structures, however, the structure and all generated potentials can be freely translated along the axis perpendicular to the surface and thus the definition of distance is more arbitrary and we discuss later the multiple definitions of adsorbate-surface distance in use. To provide a fixed reference for these potentials, we generate the potential $U_C$ for the carbon point probe with the structure initially positioned such that the uppermost surface atom defines $d'=0$, locate the distance at which $U_C(d') = 35$ \kjmol and translate the entire structure and all generated potentials by a distance $\Delta_s$ such that $U_C(d = 0.2 \mathrm{nm}) = 35$ \kjmol. This choice is largely arbitrary but provides a physically meaningful definition of the surface for amorphous or locally modified structures. The specific value is chosen to coincide with the typical value of PMFs for smooth planar surfaces at this distance.  In general, the potential is sufficiently rapidly increasing in this region that changes in the exact value of the energy chosen as a reference point produce only very minor changes in the location chosen by this procedure. For FCC (100) metal surfaces, the required translation of the structure is very close to 0, e.g. $\Delta_s = 0.007$ nm for \ce{Au} (100), while for an amorphous carbon surface (c-amorph-2) we obtain $\Delta_s = 0.17$ nm. We calculate the value which would be required for this translation for the cylindrical NPs and find it is typically on the order of $-0.03$ nm.  For consistency with the planar set, we apply this offset to the potentials initially generated with distance defined relative to the cylindrical axis and subtract $R$, such that again we have $U_C(d = 0.2 \mathrm{nm}) = 35$ \kjmol to ensure a large cylindrical NP would produce the same set of descriptive potentials as a planar NP of the same material. In Figure \ref{fig:matcompare} we plot the carbon atom probe and potassium ion probe potentials generated for three gold surfaces and three carbon nanotubes: pristine, \ce{COOH} modified (30\% by weight) and \ce{NH2+} modified (2\% by weight). As expected, the uncharged gold surfaces behave essentially identically for the two probes since these have very similar LJ parameters and differ only in terms of charge, but it can still be seen that the three surfaces themselves exhibit different interaction potentials, with the (110) surface showing a slightly wider attractive region and the (111) surface a deeper minima compared to the (100) surface. These follow from the different surface morphologies: the FCC (110) surface exhibits ridges of atoms which effectively leads to the superposition of two potentials with slightly different distances to the minima relative to a fixed surface, while the FCC (111) surface has a hexagonal structure and higher atomic density leading to a minima at approximately the same distance but of a greater depth. The CNTs, meanwhile, exhibit a stark difference between the charged and uncharged probes due to the strong charge-charge interaction present for the modified surfaces. Moreover, the high-density \ce{COOH} modification produces a clear difference in the uncharged probe as well, with the CNT surface and functional groups producing two distinct minima at different distances relative to the CNT surface. 

\subsection{Overivew of PMFs}
The PMFs consist of tabulated data consisting of pairs of values of the surface separation distance (SSD) $h$ from the nominal surface of the material to the COM of the adsorbate and the potential energy at this distance. In principle, the material surface is well-defined for materials with a smooth surface, e.g. FCC (100) crystals or planar forms of carbon, but not for materials with more complex structures, e.g. amorphous materials, FCC (110) crystals or CNTs modified with functionalised groups. Moreover, the exact choice of definition of $h$ varies between methodologies, even within PMFs supplied by the same group. During data pre-processing, four main definitions of $h$ were found to be in use: 1) the distance from the uppermost plane or CNT radius to the adsorbate COM, 2) the minimum distance between the adsorbate COM and all atoms in the material slab, and 3) the distance between the COMs of the slab and of the adsorbate minus half the total slab width, 4) the distance between the COM of the surface atoms and the COM of the adsorbate. These definitions coincide for some simple crystal structures but differ significantly for surfaces with more complex structure. In particular, definition 2) and 4) do not produce a one-to-one mapping between the vertical height above an arbitrary plane drawn through the material and the proposed definition of $h$ unless the adsorbate COM is above the highest atom. Consequently, the value of $h$ is not necessarily a simple or unique function of the distance $d$ considered in the previous sections. Indeed, even in the simplest case of an adsorbate at a distance $d$ from a plane at a reference location $d_0$, the distance $h = \sqrt{(d - d_0)^2}$ is only a single-valued function of $d$ if $d \ge d_0$ at all times. If the adsorbate is permitted to sample regions $d \le d_0$, i.e. inside the surface,  then these will be mapped to the same set of values $d$ as states outside the surface, regardless of if they represent high-energy overlaps, low-energy insertions, or a poorly defined surface plane.  By inspection of some of the input PMFs, it appears that this has occurred for at least some of the FCC (110) surfaces, which exhibit local minima or attractive states at distances which would correspond to a substantial overlap between the solid surface and the adsorbate. We have attempted to filter these out where possible to ensure the model learns only examples from which it is reasonably certain that there is a one-to-one correlation between $d$ and $h$.

To account for the fact there is potentially an arbitrary offset included in these definitions depending on the exact choice of the surface atoms, we compare the potential generated for rigid methane to the PMF for ALASCA for each of the surfaces and extract a translation distance required to move the potential onto the PMF. This is achieved by selecting the first point in the PMF with an energy under $50$ \kjmol and recording the distance of this point, then selecting the first point in the rigid methane potential with the same value of energy (i.e., $50$ \kjmol or the maximum recorded in the PMF) and recording the distance for this equivalent point. Since the rigid methane potential is defined at a known distance from the surface structure used to generate the potential, the distance $\Delta_P$ between these points defines the PMF relative to the input structure. We apply this procedure to all the surfaces describing the training set of PMFs except for three specific cases. Firstly, for \ce{CdSe} the alanine potential does not diverge at the surface as a consequence of the highly charged ions present in the structure, but since this has a smooth surface which can be assumed to coincide with the $d=0$ plane no correction is applied to these PMFs. The Au FCC (110) PMF for alanine appears improperly converged in the region $h \le 0.2$ and so for the purposes of generating this alignment we employ the equivalent Ag FCC (110) PMF, which appears to be more consistent with the others and can be expected to be a suitable proxy due to the high similarities between these surfaces. Finally, the \ce{TiO2} anatase (100) PMF does not extend to sufficiently high values of $U(h)$ and is not recorded close to the nominal surface, so in this case we perform the alignment at the lower value of $U(h) = 17.5$ \kjmol, which produces a result consistent with the other titania surfaces.  To account for other possible differences, e.g. the absence of a one-to-one mapping for SSD types 2 and 4, we also provide the SSD class obtained from the available literature for that set of PMFs as a zero-indexed categorical variable $s$, e.g. SSD class 1 has $s = 0$. For SSD class 2 we provide the distance between the uppermost atom and the set of heavy atoms assumed to form the nominal surface as it is unclear where exactly the distance is defined from in the PMFs supplied for training, while, for SSD class 4 we provide the distance from the uppermost atom to the COM of the uppermost solid layer of carbon atoms. 

\begin{figure}[tb]
    \centering
    \includegraphics[width=0.4\textwidth]{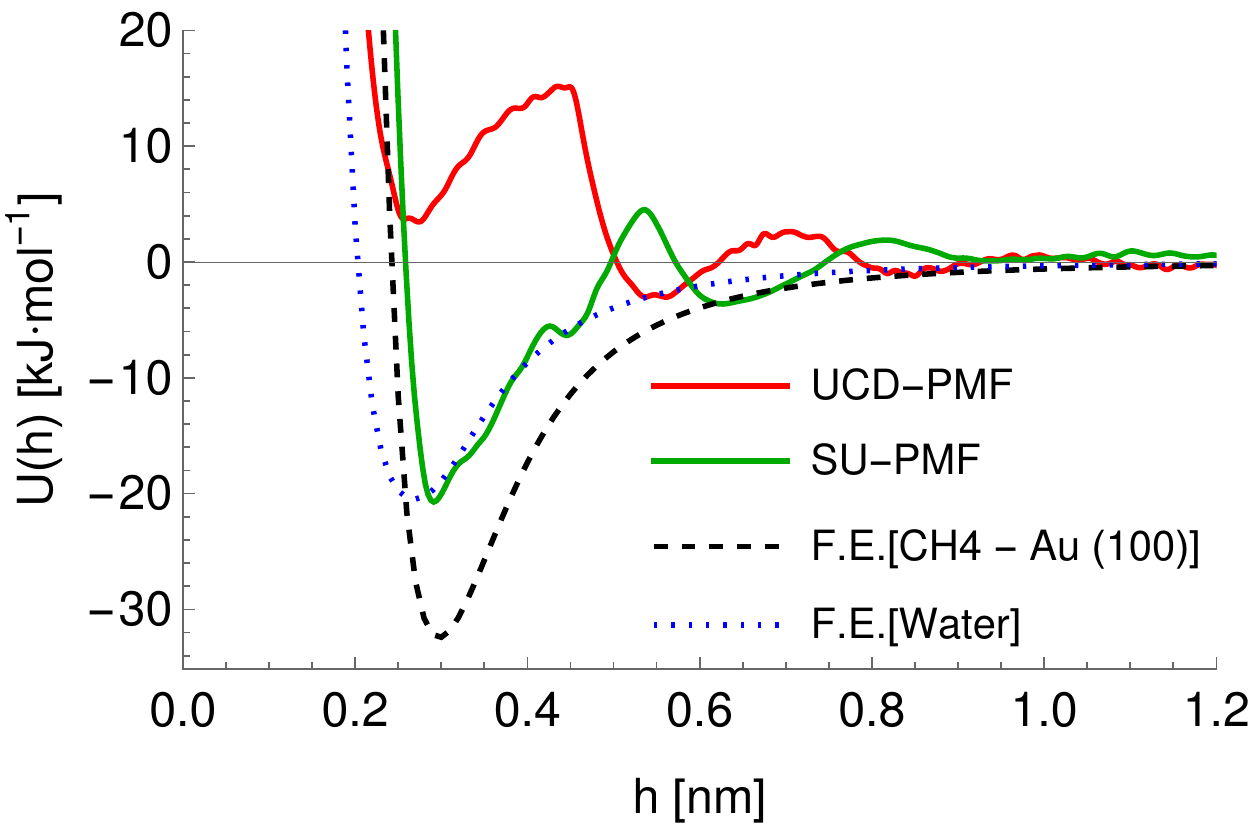}
    \caption{A comparison of potentials of molecules interacting with a gold (100) surface. Red and green lines indicate the PMFs obtained via metadynamics simulations for the alanine side chain analogue (CH4) in the UCD (red) and Stockholm (green) simulations.  The dashed black line indicates the average interaction potential for this surface with rigid methane in vacuum (GAFF model) and the blue dotted line for a single water molecule using the TIP3P model.}
    \label{fig:gold_twoversions}
\end{figure}

Further differences between methodologies also exist, which lead to different PMFs being calculated for the same system as can be seen for the ALASCA (methane) - gold (100) system in Fig. \ref{fig:gold_twoversions}. Here, the Stockholm PMF gives a very strongly binding interaction, while the UCD PMF is essentially non-binding except for a local minimum at c.a. $0.3$ nm, which is binding with respect to the next local maximum. We plot the interaction potential generated for the GAFF model of methane for both to indicate the interaction potential expected in the absence of water while treating methane as a rigid molecule, which can be seen to be a much better match to the Stockholm PMF than to the UCD PMF. We posit that the difference arise due to the use of the CHARMM forcefield in the UCD simulations in place of the GAFF forcefield employed for the Stockholm set, the inclusion of solution ions in the UCD simulations which are excluded from some (but not all) Stockholm simulations and differences in the exact type of metadynamics and criteria used for convergence and post-processing.  We observe similar effects for the remainder of the UCD (100) materials but typically no equivalent in the (110) and (111) structures, which are generally strongly binding to all chemicals. We note also that different Stockholm calculations vary in the method used to generate PMFs (MetaDF vs AWT-MetaD) and the simulation timespan, although these are expected to be reasonably consistent \cite{saeedimasine2020atomistic}. Furthermore, although both groups employ a TIP3P model of water, the SU PMFs set the $\epsilon$ parameter for the hydrogen atoms to $0$ while the UCD PMFs use a non-zero value for consistency with the CHARMM forcefield. For our purposes, we encapsulate these differences by ensuring the chemicals are described using the appropriate forcefield and by providing a categorical variable (the source variable) describing the methodology used to compute the PMF in four classes:  Stockholm-no ions, Stockholm-ions, UCD-1 (110) and (111) surfaces, UCD-2 (100) surfaces. We also include an additional variable $\Delta_H$ representing the average distance between points in the input PMF to allow the model to compensate for the different resolutions at which PMFs are recorded, which reflects possible post-processing and computational differences.

\subsection{Hypergeometric expansion of potentials}
\begin{figure}[tb]
    \centering
    \includegraphics[width=0.4\textwidth]{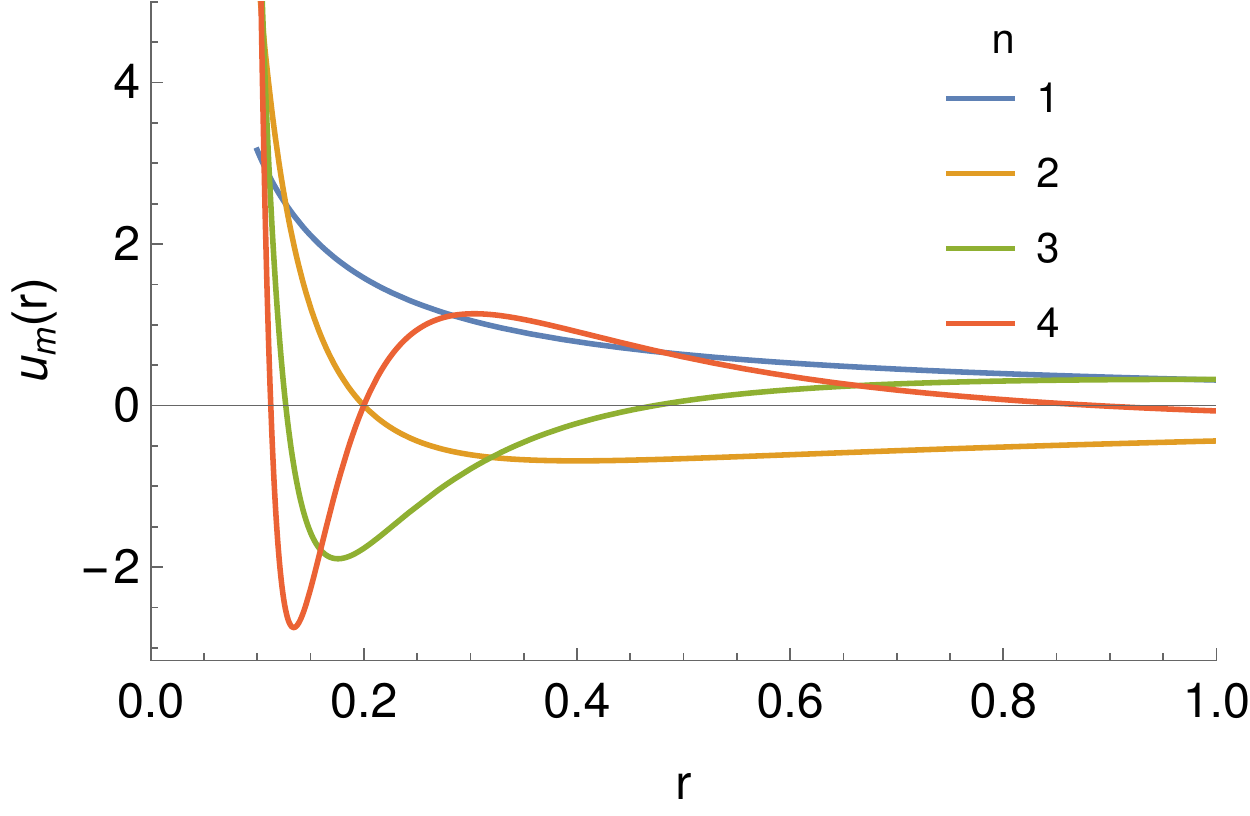}
    \caption{The basis set of functions based on hypergeometric functions used for the expansion of potentials, taking $r_0 = 0.1$ and a range of values of the parameter $n$.}
    \label{fig:hge14}
\end{figure}

The tabulated PMFs and potentials describing the materials and chemicals typically contain on the order of hundreds of pairs of distances and energies, with no consistent choice of initial and final distances, number of points per PMF, or the spatial resolution employed. Consequently, any representation of these potentials must account for all these differences while discarding as little information as possible. Directly using these paired sequences as input and output for models would greatly inflate the amount of memory required and the amount of time required to train the model, since typically sequence based methods scale unfavourably with the length of the sequence. Downsampling of the potentials risks losing valuable information, especially considering many minima are quite narrow and exist near the short-range repulsion and so may be lost by during downsampling, especially if naive averaging techniques are employed. Instead, we represent the entire potential in terms of the coefficients of a basis set expansion to preserve as much information as possible in a more compressed form.  In this way, an entire PMF can be predicted in one step, with the output providing a smooth function which can be sampled at an arbitrary resolution.  We exploit the underlying physical knowledge that the potentials represent interactions which are individually typically modelled using inverse powers of the distance $r$, e.g. the vdW potential $r^{-6}$ and the Coulomb potential $r^{-1}$, and that the potentials obey similar boundary conditions, i.e. they diverge for $r \rightarrow 0$ and tend towards a constant which may be set equal to zero for $r \rightarrow \infty$.  To take advantage of this, we apply the Gram-Schmidt orthonormalisation method to the set of functions $1/r^i$ to construct an orthonormal basis set of functions $u_m(r)$,
\begin{equation}
u_m(r) = \sum_{i=1}^{i=m} c_{m,i} r^{-i}
\end{equation}
with the definition of orthonormality given by
\begin{equation} \label{eq:orthonormalEq}
    \int_{r_{0}}^\infty u_m(r) u_n(r) dr = \delta_{mn},
\end{equation}
where $\delta_{ij}$ is the Kronecker delta function which is equal to unity for $m=n$ and is otherwise zero. Here, we have chosen the functions to be orthonormal with respect to an inner product defined by integration over the interval $[r_0,\infty]$, where $r_0 > 0$ is used to avoid divergence at $r=0$. We assume that $r_0$ is equal for all terms in a given expansion and discuss its selection later. The required coefficients $c_{m,i}$ are functions of $r_0$ and can be found by solving the set of algebraic equations obtained by evaluating Eq.~\eqref{eq:orthonormalEq} for successive values of $m,n$. Using Wolfram Mathematica \cite{Mathematica13p1} to calculate the coefficients for $m \le 20$ and finding a closed-form solution valid in this region, we have been able to empirically determine that for at least up to the $m=20$ term the resulting series may be conveniently expressed in terms of a hypergeometric (HG) function \cite{dlmf} $_2F_1(a,b,c,r)$
\begin{equation}
    u_m(r) =(-1)^{m+1} \sqrt{2 m-1} \frac{ \sqrt{r_0} }{r} \, _2F_1\left(1-m,m;1;\frac{r_0}{r}\right) ,
\end{equation}
or equivalently in terms of a sum,
\begin{equation}
    u_m(r) = - \sum_{i=1}^m \frac{P_H(1-m,i-1)  P_H(m,i-1) }{\left(P_H(1,i-1) \right){}^2}  \sqrt{2 m-1} (-1)^{m} r^{-i} r_0^{i-\frac{1}{2}} ,
\end{equation}
where $P_H(m,i)$ indicates the Pochhammer symbol conventionally denoted $ (m)_i$  \cite{dlmf}.  We have numerically confirmed these functions possess the required property of orthonormality over the interval  $[r_0,\infty]$ for $m \leq 20$. We plot $u_m(r;r_0)$ for a range of values of $m$ in Figure \ref{fig:hge14} to illustrate their general form. A function of interest $U(r)$ can be expanded in terms of these functions,
\begin{equation}
    U(r) = \sum_{m=1}^{m=M} A_m u_m(r),
\end{equation}
where the property of orthonormality can be used to express the required $A_m$ coefficients by,
 \begin{equation}
    A_m = \int_{r_0}^\infty u_m(r) U(r)
 \end{equation}
provided that the same value of $r_0$ is used for all functions in a given expansion. For our purposes, the expansion is performed numerically for the tabulated potential or PMF up to the highest required order of $m$ once a value of $r_0$ has been selected, taking $r=d$ or $r=h$ as needed and denoting the expansion parameter $r_0$ for both for historical reasons and consistency with the code. We truncate the expansion after the $m=20$ term, finding this generally gives good results even for PMFs with very sharp features. To generate the data sets for use later, we perform the expansion for PMFs at a range of values of $r_0$ in the range $0.1 $to $ 1.0$ nm and for potential probes at $r_0 = 0.2$ nm. Before expansion, potentials are shifted such that $U(r_{\mathrm{max}}) = 0$, taking $r_{\mathrm{max}} = 1.5$ nm.  Generally, this leads to an insignificant except for the interaction potentials between two charged species which decay much more slowly and so have non-zero values at $r_{\mathrm{max}}$.

 \subsection{ANN model for PMF prediction}
 In the previous sections, we have provided methodology for representing materials and chemicals of interest in terms of a set of coefficients describing their interaction potentials with probe molecules. These coefficient sets are similar to the feature maps produced by classification and image recognition neural networks, e.g. ResNet, while the remapping of the input potentials to an output potential is conceptually similar to translation and style transfer techniques implemented in various language and image models, again employing neural networks \cite{pml1Book}. Taking inspiration from this, we employ a neural network method and treat the set of chemical and surface potentials as a feature map and use an encoder approach to reduce this to a low-dimensional space employing both convolutional and fully-connected layers to produce multiple representations of the system. This encoder serves as an initial feature selection to determine the most relevant parts of the input data for the prediction of the PMF. These are combined with estimates of the energy at $r_0$ and the minimum energy in the region $h>r_0$ to produce a final encoded state, to which the inputs describing the particular PMF ($r_0$, surface offset $\Delta_s$, PMF offset $ \Delta_P$, resolution $\Delta_H$, the source, SSD, and offset variables) are appended. The $E(r_0), E_{min}$ estimates are generated using fully-connected networks operating on values obtained from the set of input potentials including their minima and values at a fixed reference point and the categorical variables. These are sent directly as output for optimisation, and copies with back-propagation blocked passed as input to the remainder of the network. To accelerate the training of the remainder of the network, we use a stochastic teacher forcing approach in which the true value is sent $50\%$ of the time and the predicted value the remainder of the time, in either case applying noise and normalisation to produce a form useful for input.  The PMF itself is generated through a multi-step approach. First, a set of expansion coefficients is generated using a small fully-connected network directly from the encoded state. During the development of the model, this was found to not produce a sufficiently accurate PMF and so it requires further refinement. We generate additional potentials by mixing together the input potentials, using the encoded state as the input to a small set of fully-connected layers to produce the mixing coefficients, one for each input potential. The weighted sum of the current set of potentials is passed through a non-linear activation layer to produce a new potential, which is appended to the list of known potentials. This procedure is repeated a small number of times with the newly created potentials appended to the list of input potentials to allow for the generation of more complex potentials. Additional potentials are generated by a convolution-transpose network from the encoded state, which starts from an initial set of three coefficients and up-scales these to take into account the sharing of information between neighbouring coefficients in a given potential. Another set of mixed potentials is generated from all these candidates and from this the initial output potential is generated. This output potential is then refined by multiplication by a matrix generated with coefficients computed from the input parameters independent of the structures in question, i.e., on $r_0$ and the set of categorical variables. This final step is done to provide any necessary translation or transformation of the potential which is independent of the exact chemical identity. As is typical for neural networks, the number of free parameters is substantially higher than the number of data points and so we employ regularisation techniques, primarily dropout and Gaussian noise (both additive and multiplicative) to reduce the risk of overfitting \cite{pml1Book}.
 
 Throughout the network, we generally use residual connections to accelerate learning and allow for additional layers to be added without decreasing the performance of the model \cite{pml1Book}. The coefficients describing the input potentials take both positive and negative values and vary over multiple orders of magnitude and we require that the model remains sensitive to small changes in coefficients with absolute values close to zero without clipping large values. We therefore employ an activation function of the form,
 \begin{equation}
     f(x) =b_2 + \mathrm{sgn}(x + b_1) \alpha  \mathrm{log}( |  (x + b_1)/\alpha |+  1)
 \end{equation}
 where $\alpha,b_1,b_2$ are parameters learnt individually for each activation during training and $\mathrm{sgn}(x)$ is the sign function equal to $+1$ for $x \ge 0$ and $-1$ else. This function behaves similarly to the traditional sigmoid activation function but does not saturate to a constant value for large absolute values of $x$ and instead logarithmically diverges, while ensuring the sign of the input is maintained to differentiate between positive and negative inputs. The parameter $\alpha$ controls the rate at which the function moves from linear to logarithmic behaviour, with $\alpha \rightarrow \infty$ producing a linear activation and the limiting behaviour $\alpha \rightarrow 0$ producing $f(x) = 0$. The two bias variables $b_1,b_2$ enable translation of the input and output respectively to increase the flexibility of this function.  We initialise $\alpha$ using the Glorot normal initialiser implemented in Keras and the bias values to small constants close to $0$.
 
The primary input is a value of $r_0$ for the PMF and a set of HGE coefficients describing the interaction potentials of the chemicals and surfaces with a set of probes as discussed in Section \ref{section:inputDefinition}. All input potentials are described using an expansion value of $r_{p,0} = 0.2$, which typically ensures that the entire region in which the potential is attractive is included. The value of these potentials at this point and the global minimum for each is passed as further input.  We normalise $r_0$ by transforming this to the log-domain and rescaling the resulting variable to have mean and variance of $0$ and $1$ respectively and pass this as a model input. The log-transformation is done to ensure that the model is sensitive to small variations in $r_0$ at small values of this parameter, which significantly change the expansion coefficients.  We assign further variables to account for differences in geometry and methodology used to generate the PMFs as discussed previously. In generally, these categorical variables are encoded in the datasets using an integer and converted to a one-shot encoding by a pre-processing layer, then mapped from $(0,1)$ to $(-0.5,0.5)$. To ensure the network does not over-specialise to one particular category, a noise layer randomly perturbs these encodings during training by multiplying them by $-1$ with a probability of $0.1$ and applying Gaussian noise.  In total, the categorical values consist of the source variable defining general simulation properties, a second defining the shape (planar, cylindrical) and a third defining the convention used for the SSD (upper surface - COM, minimum atom-COM distance, adjusted centre slab - COM distance, surface COM - COM distance). For most input potentials, we provide only the free energy averages, with the potentials obtained from the minimum energy at each value of $r$ used for the carbon atom, water and carbon ring potentials for both surfaces and chemicals, with methane-minimum additionally provided for surfaces. The goal of providing these minimum energy potentials is to enable the model to distinguish between a surface with high and low regions of binding affinity and a uniform surface of medium binding affinity, since both of these may have similar average free-energy potentials. Likewise, for chemicals this enables the model to learn differences between isotropic molecules and those with regions of high and low binding affinity or hydrophobicity.

To train the network, we employ the Adam optimiser with empirically adjusted learning rate and $\epsilon$ parameter \cite{kingma2014adam} for 50 epochs. During development of the model, we have explored a number of loss functions to overcome the issue of the different characteristic magnitudes of the various outputs $(A_i, E_0, E_{min})$ and ensure the model does not specialise to one of these at the expense of the remaining outputs. In particular, we find that root-mean-square and related loss functions (Huber loss, MSE, absolute error) tend to over-emphasise the lower-order coefficients, primarily $A_1$, and only gradually fit the higher-order coefficients.  A further issue is the fact that the root-mean-square deviation between a predicted and target PMF as averaged over the entire PMF is weighted more strongly towards the $h \rightarrow 0$ region in which the potential diverges towards positive infinity. In this region, a slight horizontal displacement of the PMF corresponds to a very large, but physically meaningless error, since large positive values correspond to essentially a zero probability for the adsorbate to be located there. The relative error, meanwhile, diverges for values of the potential close to $0$ and so a loss function based on this value over-emphasises the long-range section, which again is of less physical interest. To overcome these issues, we define a loss function based on the Kullback-Leibler (KL) divergence \cite{pml1Book}, which measures the distance between a target probability density $p(x)$ and an approximate one $q(x)$,
\begin{equation}
    \mathrm{KL}[ p(x),q(x)] = \int  p(x) \ln{ \left[\frac {p(x)}{q(x)} \right]} dx. 
\end{equation}
The PMF is related to the probability density $f(h)$ for the position of a particle in that potential, $f(h) = B \exp{[-U(h)/k_B T]}$, where $B$ is a normalisation constant to ensure the total probability integrated over the interval $[0,\delta_c]$ is equal to unity,
\begin{equation}
    B^{-1} = \int_{r_0}^{\delta_c}   \exp{\left[-U(h)/k_B T\right]} dh  =  \delta_c \exp{\left[  -E_{ads}/(k_B T) \right]} ,
\end{equation}
where we assume $U(h \le r_0) = \infty$ and take $\delta_c = 1.5$ nm and $k_B T = 1$. Taking $p(h)$ to be the density for the target PMF, $q(h)$ the density for the predicted PMF, and using the definition of $B$ in terms of the adsorption energy we find
\begin{equation}
    \mathrm{KL}\left[ U(h), \hat{U}(h) \right] = \frac{E_a - \hat{E}_a}{k_B T} + \frac{ 1}{k_B T \delta_c} \int_{r_0}^{\delta_c} e^{E_a-U(h)/k_B T} \left[ \hat{U}(h) - U(h)\right]  d h.
\end{equation}
The first term in the above expression is simply the difference between the adsorption energy for the target and predicted PMFs, while the latter is a measure of the difference between the PMFs themselves with a weighting function $\exp{\left[E_a - U(h)\right]}$ applied. This weighting function is smaller where the target PMF takes large positive values and greater where the PMF is large and negative, reaching its peak in the most strongly binding regions.  Thus, minimising this loss function helps to ensure that the PMF is most accurate in the physically relevant regions. The choice of $k_B T = 1$ is used to ensure that the loss function remains relatively sharp towards minima, as during initial testing it was found that using a larger value corresponding to a physical temperature of $T = 300$ K reduced the accuracy of the model. We evaluate this loss numerically by sampling the PMFs generated for the target and predicted coefficients on a grid and approximating the integration by summation over these points. Formally, the KL loss is non-negative for all input functions and so can be directly used as a loss function by summation of this value over all PMFs in a batch, but in practice we use the mean-square value calculated over a batch to stabilise the training for values of the loss close to zero. We combine this loss with the mean squared error for each of the $A_i$ and the two values $E_0, E_{min}$, weighting each of these by the variance for that output variable estimated from the training set to ensure that the optimisation treats each of these equally. Without this weighting, we find the training emphasises the fitting of the $A_1$ parameter which typically varies over the widest range of values and thus has the largest mean-square error but controls only the coarse long-range behaviour of the output potentials. To counteract the unbalanced nature of the dataset, e.g. the high proportion of PMFs for planar surfaces compared to cylindrical surfaces, we assign sample weights to each PMF which are used during the evaluation of the loss function to ensure PMFs with rarer features contribute more to the training and avoid biasing the network towards the most common PMFs. These sample weights are generated based on several criteria. Each categorical variable contributes a factor to the weight proportional to the inverse of the frequency of that value, such that PMFs consisting of a category with few examples are weighted more strongly. The AGGLOMERATE clustering algorithm implemented in scikit-learn \cite{scikit-learn} is used to assign sets of $A_i$ values to clusters in order to identify PMFs with dissimilar features and weights are assigned based on the inverse frequencies of these. Finally, we also include a factor derived from the minimum of the PMF in the region $h \ge r_0$, normalised by the mean and standard deviation of the PMF minima in the entire training set. This ensures that the training algorithm weights especially strongly and weakly binding PMFs more in order to reproduce the correct behaviour at both extremes.

\subsection{Generation of training data}
For training purposes, it is advantageous to provide multiple sets of $A_i$ values for each PMF and set of input potentials to increase the size of the dataset which is otherwise fairly limited. We therefore apply the HGE procedure to each PMF taking  multiple values of $r_0$ for each PMF to generate multiple sets of coefficients. This has the additional benefit of dealing with the issue that the PMFs are generally truncated at different minimum values of $h$ by providing examples of the same PMF with different degrees of truncation, and further eliminates the need to choose a specific value for $r_0$. During the expansion, we record both the actual value of $U(r_0)$ and the value obtained from the generated expansion, $U_{H}(r_0)$. We discard results where $|U_(r_0) - U_H(r_0)|^2 \ge 10$ or where the error increases with increasing $m$ since this indicates that the expansion has failed to converge. For each PMF we record the methane-to-alanine offset distance $\Delta_P$ calculated for the surface in question as an additional parameter to allow the model to compensate for the unknown location of the surface used in the definition of the PMF. To increase the size of the training set we generate additional expansions in which we translate the PMFs by $ \Delta_P$ and a random value drawn from a zero-mean normal distribution with standard deviation $\sigma = 0.1$, again recording the final offset relative to the methane potential. Consequently, the initial PMF has a recorded offset of $\Delta_P$ while translated PMFs have an offset typically in the interval $(-0.1, 0.1)$. These translated PMFs are further perturbed by small amounts of noise prior to applying the HGE for a given value of $r_0$. Denoting a normal distribution of mean $\mu$ and standard deviation $\sigma$ by $\mathcal{N}(\mu,\sigma)$, these perturbations are random translations of the entire PMF by an additional small random amount $\delta_h \sim \mathcal{N}(0, 0.05)$, multiplication of the energy values for the entire PMF by $\alpha \sim \mathcal{N}(1,0.1)$ and the application of a small amount of additive noise $\eta_i \sim \mathcal{N}(0,0.2)$ to each individual energy value in the PMF such that the noisy PMF is given by
\begin{equation}
    \tilde{U}_p(h_i) = \alpha U_p( h_i + \delta_h) + \eta_i.
\end{equation}
This procedure has the advantage of smoothing out some of the noise inherent in each PMF and reducing the risk of the neral network overfitting to the specific examples provided, and is essentially the one-dimensional equivalent of the typical image augmentation techniques of adding shot noise, randomly translating the images, and randomly adjusting the brightness of the entire image, all of which are known to improve both the training and validation of networks \cite{pml1Book}.   Multiplication by $\alpha$ maps directly onto the coefficients of the modified PMF, $\tilde{A_i} = \alpha A_i$, but the modifications in the perturbed coefficients due to translation or shot noise are much more difficult to express in terms of a simple transformation of the HGE coefficients.   Thus, these transformations are pre-applied to the PMFs before expansion to generate four noise replicates at each value of $r_0$ for each PMF. Compared to implementing these noise transformations directly in the network, this method has the benefit of producing a larger dataset to optimise over and thus smoothing out the loss function for an individual epoch. The downside of this method is that it increases the memory required for the dataset and does not produce a new random sample for each epoch. Thus, we also apply similar noise directly in the HGE domain implemented as Tensorflow layers for the sets of input potentials to provide new perturbations for every training epoch without an increase in the memory required for the training set while providing some protection against overfitting. Noise is additionally applied to all outputs as a form of label smoothing, again to reduce overfitting.

Due to the limited size of the available dataset, the majority of available PMFs were used for model development with a small number reserved for final testing. Some PMFs have been manually excluded due to being clear outliers for reasons which could not be resolved during model development, namely: AFUC, TRPSCA, PHESCA for \ce{Au} (100) UCD; ALASCA, CYSSCA, LYSSCA, HIESCA, for \ce{Au} (110) UCD; GLUSCA, BGLCNA and TYRSCA for \ce{Ag} (100) UCD; ASPPSCA, LYSSCA and THRSCA  for \ce{Ag} (110). Typically, these exhibit either spurious maxima or minima, or appear to have allowed penetration of the molecule past the nominal surface. Predictions are still made for these PMFs and they are not excluded from the calculation of final train and test statistics. The \ce{Ag} (100) and \ce{Au} (100) PMFs are set to a separate methodology (UCD-2) due to their clear difference from the remaining UCD FCC metal PMFs, but otherwise treated normally. For the remaining PMFs, we have tested two variants of generating the training and validation sets. In the first variant,  the sets of materials and chemicals (identified by SMILES code) are individually split into training and validation sets, with any PMF featuring a validation material or chemical excluded from the dataset used for training the model. This produces a total of four classes of PMF: training material -- training chemical, training material -- validation chemical, validation material -- training chemical, and validation material -- validation chemical. We manually assign the gold FCC (100) structure from both sources to the training set in order to provide the model with a comparison between the two sources for the same structure. Since the Stockholm-sourced gold PMFs typically exhibit a very strong binding energy, this has the further benefit of ensuring the model is valid over a wide range of interaction strengths. Likewise, the gold FCC (111) structure is manually assigned to the training set since this exhibits an even stronger binding energy and to ensure that the class of non-(100) UCD PMFs is represented. To generate the rest of the training set, we employ a clustering algorithm to identify broad classes of surfaces and chemicals and ensure the training set contains examples of all classes. To do so, we use the AGGLOMERATE clustering algorithm implemented in scikit-learn \cite{scikit-learn} based on the coefficients describing the input potentials up to $8^{\mathrm{th}}$ order, with a maximum of fifteen clusters allowed each for chemicals and materials. A randomly selected example from each cluster is assigned to the training set such that this covers as wide a variety of surfaces and chemicals as possible. The remainder of the training sets are chosen from all remaining materials and chemicals. For the materials, we generally find many clusters consist of a single example (e.g. \ce{CdSe}, \ce{Fe2O3}) while a large cluster contains almost all the CNTs. Depending on the exact parameters chosen, the FCC metals are either combined into a single cluster or separated into a group containing (100),(111) surfaces and a second containing the (110) as a consequence of the roughness of the (110) surface compared to the other two. We train five variants of the model using different random splits generated in this way and demonstrate later that this produces acceptable results in most cases, but the set for which both material and chemical are excluded from the training set typically exhibits a number of mis-predictions. Thus, to produce reliable predictions and maximise the range of materials for which reliable predictions can be made, we employ a bootstrap aggregation (bagging) method. In this method, we again train ten variants of the model for fifty epochs each using the same architecture but each using a different dataset selected by random resampling from the set of all PMFs, with training and validation sets selected from these resampled sets at random.  This bootstrap aggregation procedure is known to produce more reliable predictions in general and potentials in particular \cite{agrafiotis2002use, schran2021machine}, ensures that there is a non-zero probability for all PMFs in the dataset to contribute to the final model, and has the additional benefit of providing estimates of the uncertainity of each prediction by comparing the output of each model. Typically, circa $500$ distinct PMFs contribute to a given bootstrap replicate after the resampling and training-validation split. For reproducible results, the seed values used for the random number generation in Python, NumPy and Tensorflow are fixed based on the model type.

For final testing and comparison of the bootstrap ensemble, we use surfaces not employed in the development of the model (copper, iron and an additional amorphous carbon structure) and two additional small molecules calculated in the UCD set (beta-galactose, choline). The former is designed to assess the ability of the model to make predictions of the PMFs required for the operation of the UnitedAtom adsorption model\cite{power2019multiscale} for new materials and is a key outcome of this work. The prediction of adsorption profiles for small molecules is of interest but due to the small amount of available data only limited testing of this functionality can be achieved here. Testing results reported were evaluated using the version of the model archived at \cite{rousepmfparchive}; all results shown in this work correspond to the model trained before prediction was performed for this test set. Full details of which potentials are supplied as input are provided in the model repository, as is the Python code used to generate the network and details on the training-validation splits. 

\subsection{Implementation}
All scripts are implemented in Python 3 using primarily the NumPy, SciPy, Pandas, Tensorflow, scikit-learn libraries \cite{tensorflow2015-whitepaper,scikit-learn}. Calculation of potentials and training of the neural network are performed on a Dell Precision 7910 workstation with a Xeon CPU E5-2640 v4 running at 2.40GHz. A training epoch on the noise-augmented dataset takes on the order of 20 to 30 minutes utilising the CPU only. Parameterising a new chemical takes on the order of a few minutes while each surface takes up to a few hours for one CPU core for the set of point probes, plus an extra hour for each molecular probe. Optimisation of this bottleneck remains a future goal, but we note that multiple surfaces may be parameterised simultaneously and that this remains substantially faster than direct computation of PMFs.

\section{Results}
To demonstrate a typical output of the procedure for generating a numerical representation of a molecule, the low order coefficients for a set of probes to the tryptophan side chain analogue are presented in Figure \ref{fig:ljchempointslab}. Potentials for the full set of surfaces and chemicals and the HG expansion coefficients corresponding to the results presented here are archived at \cite{rousepmfparchive}. This repository also contains tabulated binding energies for all predicted PMFs using both ensemble methods. The code repository \cite{Rouse_PMFPredictor-Toolkit_2022} contains the most recent values reflecting any changes in the code or addition of new molecules or surfaces. All results in this section are obtained from the model checkpoints with lowest training loss, for which the validation loss is typically also a minimum.

 \begin{figure}[tb]
    \centering
    \includegraphics[width=0.4\textwidth]{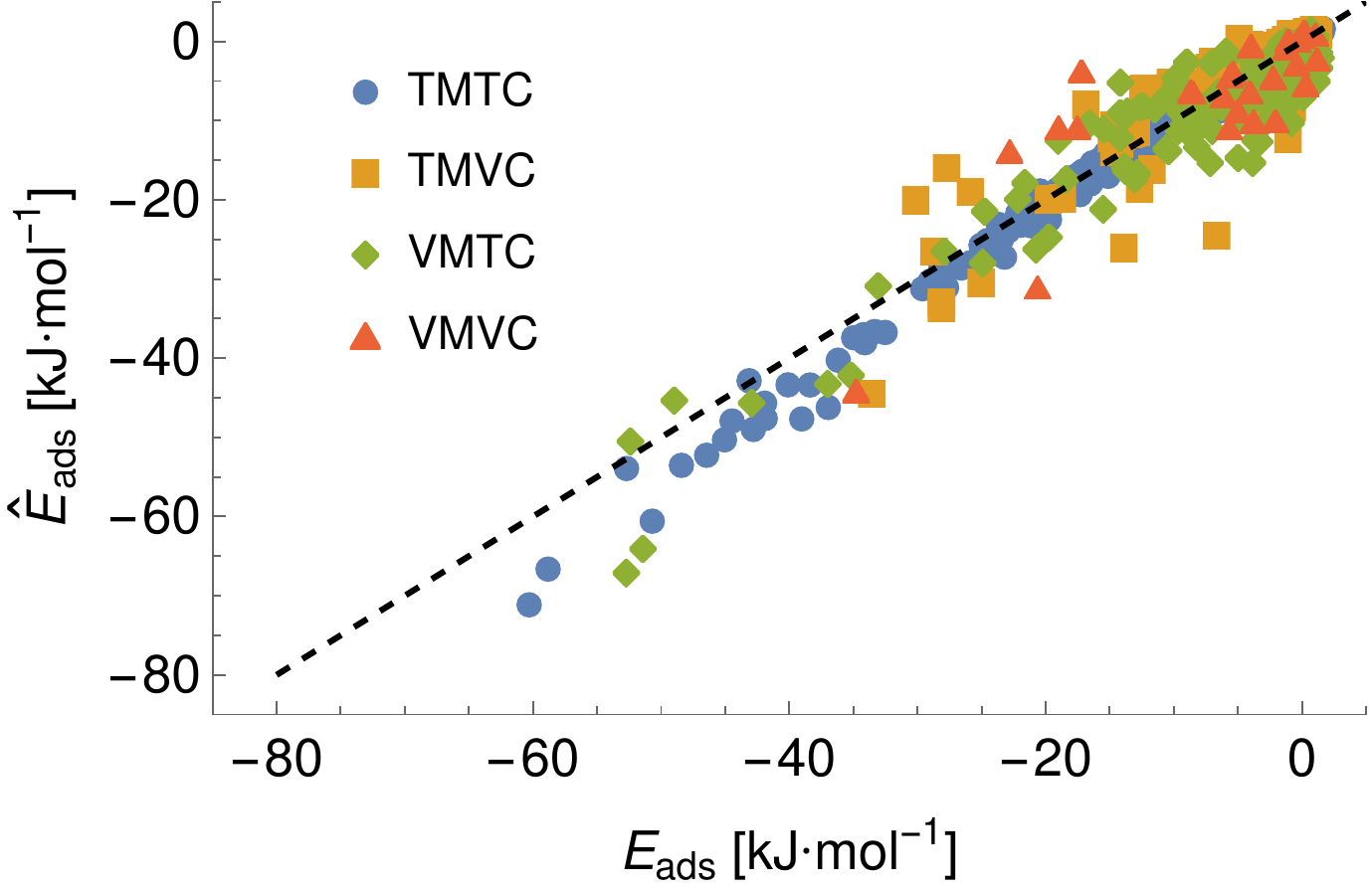}
    \caption{A comparison of the predicted adsorption energies to those extracted from the input potentials of mean force, generated using model ID Cluster-A-1 and showing the four prediction classes arising from combinations of Training and Validation sets of Material and Chemicals, e.g.  Training Material--Training Chemical (TMTC). The black dashed line indicates the ideal case of exact prediction and is shown to separate the regions of over- and under-prediction. }
    \label{fig:eadspredicted}
\end{figure}

\begin{figure}[tb]
    \centering
    \includegraphics[width=0.4\textwidth]{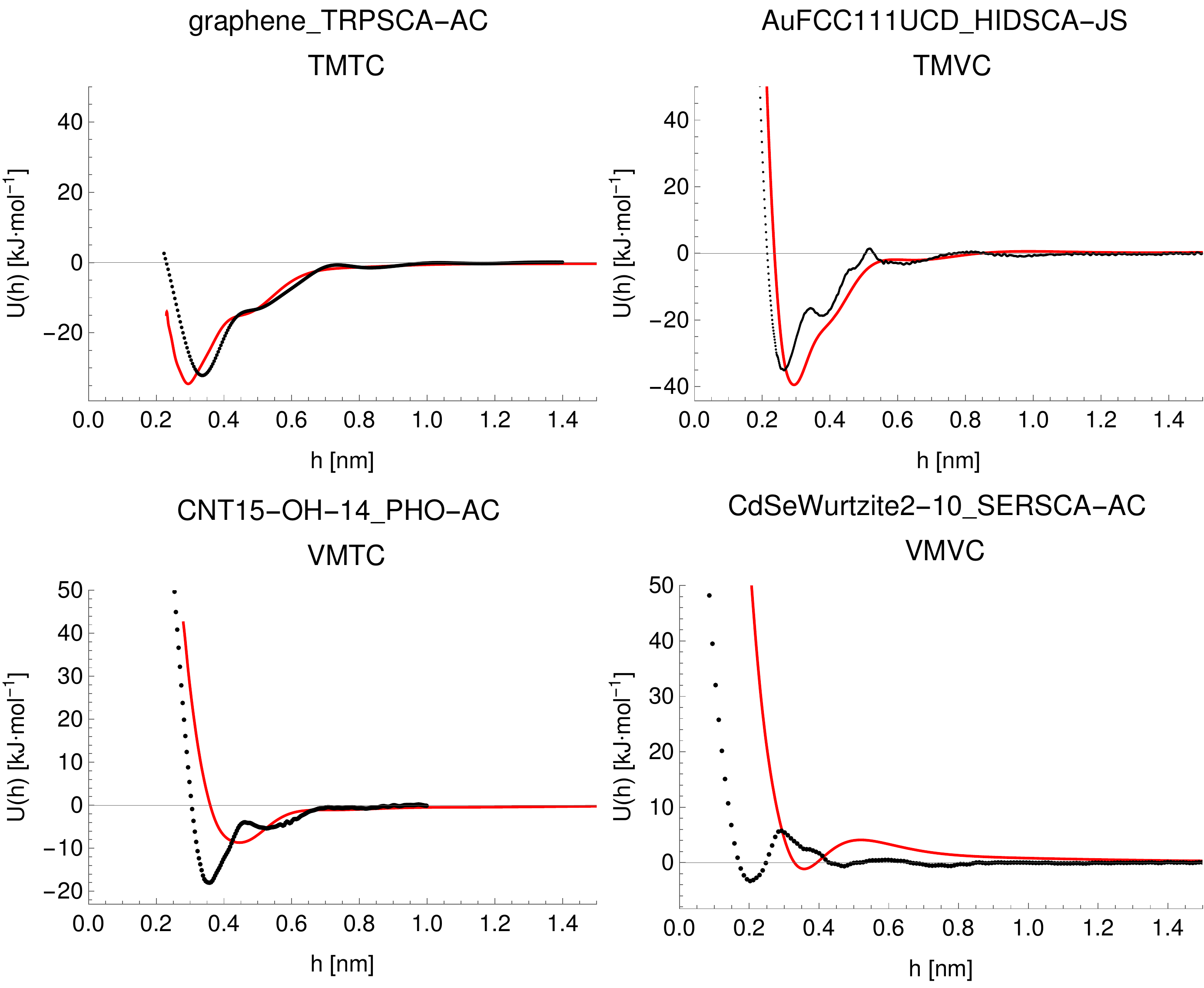}
    \caption{Predicted (red) and input (black) PMFs with the highest KL divergences for each of the four classes of training/validation material/chemical pairs for model ID Cluster-A-1, with PMFs predicted with parameters matched to the original inputs to obtain a prediction using the same conventions as the original. }
    \label{fig:kldivgraph}
\end{figure}

Given the large number of predicted PMFs and models, here we only present some examples and summary statistics, with the full set of predictions available for download at \cite{rousepmfparchive} and results for selected materials and the testing chemicals available in the ESI. For each model variant, we compute the PMFs, KL divergences, and binding energies at $T=300 \mathrm{K}$ for all material--adsorbate pairs, matching simulation type and SSD parameters to the ones used for training the model and taking $r_0$ to be as close to $0.2$ as possible based on the input PMF. The results for one example model (cluster split, no bootstrapping, split ID $1$) are plotted in Figure \ref{fig:eadspredicted} in comparison to the binding energies calculated for the original PMFs with the worst-performing predictions for each of the four classes in terms of KL divergences shown in Figure \ref{fig:kldivgraph}. The agreement is generally quite good and we find a high correlation for both training and validation groups for this model. The poorly-performing PMFs can be seen to generally exhibit the correct structure aside from the PMF for phosphate binding to an \ce{OH} modified CNT (VMTC class) capturing only the second, weaker adsorption and the horizontal translation of the \ce{CdSe} - serine SCA PMF. This latter case is attributed to the fact that \ce{CdSe} is in general non-binding and the same general structure is observed for the majority of the other PMFs for this surface, with SERSCA providing one of the few exceptions. This highlights the importance of employing a diverse training set to capture such outliers.  In the ESI we give summary statistics for the binding energy and KL divergence for groups of networks employing different random seeds and both split methodologies. We observe that there is a high degree of variability in the accuracy of the predictions for the cluster-based random splitting when attempting to predict materials from the validation set, with some splits producing very good results on unseen materials and others failing to converge. Two of the cluster-based splitting models (Cluster-A-3, Cluster-A-4) perform significantly worse on validation data and so are not used for further study. This likely relates to the highly heterogeneous nature of the set of materials, which may require further refinement of the cluster-based assignment of outliers to the training set. The bootstrap replicates generally exhibit more reliable validation scores (see Table \ref{tab:EAdsStatsAvg} and Table S5 in the ESI) and so we recommend the use of results from the ensemble of these, but provide results for both ensembles excluding the two poorly performing cluster models. 

\begin{table}[b]
    \centering
\begin{tabular}{ccccc}
\hline
 \text{} & \text{TMTC} & \text{TMNC} & \text{NMTC} & \text{NMNC} \\
 \hline
 \text{Bootstrap correlation } & 0.96 & 0.59 & 0.89 & 0.55 \\
 \text{Bootstrap $R^2$ } & 0.92 & 0.35 & 0.8 & 0.3 \\
 \text{Cluster correlation } & 0.97 & 0.53 & 0.89 & 0.54 \\
 \text{Cluster $R^2$ } & 0.93 & 0.28 & 0.79 & 0.29 \\
\end{tabular}
    \caption{ Summary statistics for the accuracy of binding energies (correlation and $R^2$) for the ensemble averages for the bootstrap and cluster split methods. In the class labels, T and N refer to training and novel, M and C to material and chemical, where training species were present at least once in the training set.}
    \label{tab:EAdsStatsAvg}
\end{table}

\begin{figure}[tb]
     \centering
     \includegraphics[width=0.4\textwidth]{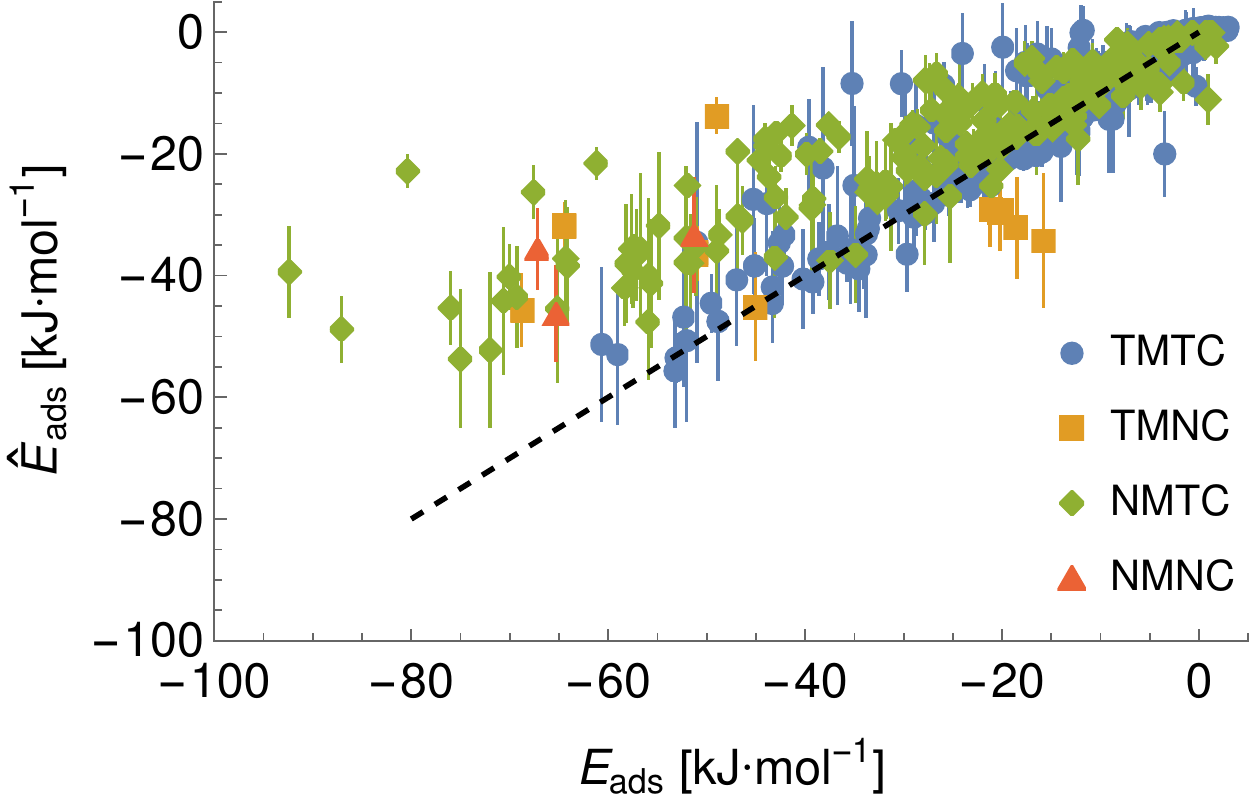}
     \caption{Binding energies predicted by the final bootstrap ensemble. Binding energies are extracted from the linear average PMF and compared to the values predicted via metadynamics. Here, TMTC indicates training material - training chemical, NMTC indicates novel material - training chemical and so on. Error bars indicate the standard deviation of the adsorption energy set extracted from individually predicted PMFs while points give the energy extracted from the average PMF.}
     \label{fig:noveleads-bootstrap}
 \end{figure}

Final prediction of PMFs is achieved by averaging the PMFs generated from all ensemble members, which implicitly allows for all material-chemical pairs to feature in the training set. Thus, to validate this model, we must employ the (limited) data not otherwise used in the model development process. To test the ability to make predictions for new chemicals, we calculate PMFs for two extra molecules: 2-acetyl-2-deoxy-beta-d-galactoseamine (BGALNA) and choline (CHOL), which are compared to the predicted PMFs generated for the FCC Au and Ag surfaces (UCD methodology, three surface indices, see ESI Figure S2 for BGALNA). Predictions for new surfaces are performed for an alternate amorphous carbon morphology c-amorph-3 and additional FCC metals Cu \cite{js_copper} and Fe \cite{parinaziron2022}. We match the input parameters for prediction to those assumed for the style of PMF but do not calculate the alanine offset. The generated PMFs and adsorption energies are provided in the repository \cite{rousepmfparchive} and binding energies extracted from these presented in Figure \ref{fig:noveleads-bootstrap}, with adsorption energies for the novel chemicals to FCC metals listed in the ESI Table S6, energies for the training chemicals to the amorphous carbon surface in Table S7, and to Cu (111) in Table S8. We find a generally correct ranking for novel surfaces but worse agreement for novel chemicals. We attribute this discrepancy  to the limited amount of data for novel chemicals and the noise and inconsistency in the target PMFs. In particular, BGALNA should be similar to BGLCNA but is found to differ significantly in metadynamics across surfaces despite the similar surface input potentials (ESI Figure S2). CHOL is typically predicted reasonably accurately except to (111) faces, for which the predicted binding energy is much more favourable than that found through metadynamics, e.g. -34 vs -16 \kjmol for Ag (111). The reason for this is unknown but appears consistent across the models.

   \begin{figure}[tb]
     \centering
     \includegraphics[width=0.4\textwidth]{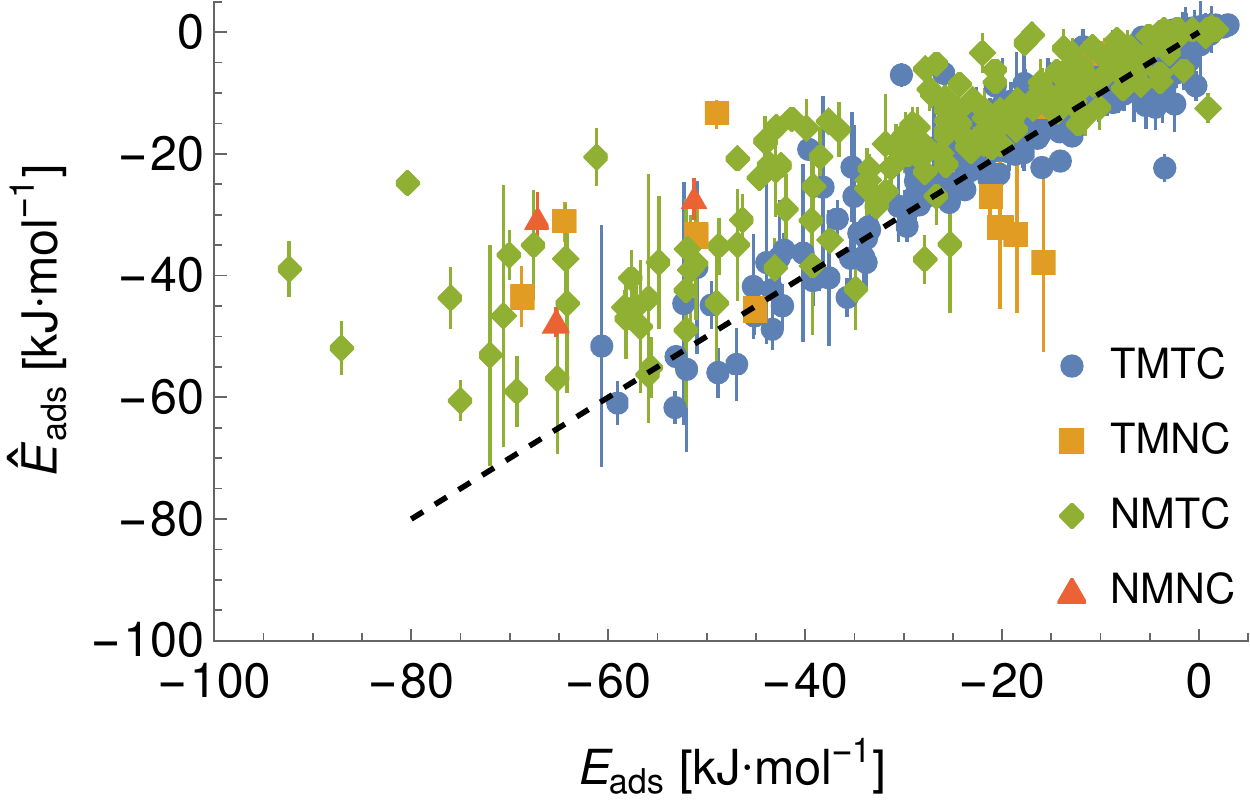}
     \caption{Binding energies predicted by an average over the three best performing cluster split models. Binding energies are extracted from the linear average PMF and compared to the values predicted via metadynamics. Here, TMTC indicates training material - training chemical, NMTC indicates novel material - training chemical and so on. Error bars indicate the standard deviation of the adsorption energy set extracted from individually predicted PMFs while points give the energy extracted from the average PMF.}
     \label{fig:noveleads-cluster}
 \end{figure}

The methodology proposed here uses descriptors for the surfaces and chemicals which can be derived directly from their structure and forcefield parameters and so is conveniently extendable to new structures for both. To demonstrate the power of this approach, we use the trained model to make predictions for approximately one hundred additional small molecules taken from the ChemSpider database, making the selection based on those consisting of the ``standard'' elements for organic molecules and with total molecular mass of under 200 AMU, selecting the hundred most highly cited and discarding those already present in the dataset. Details of the full set are available in the ESI, where we report the ChemSpider ID, SMILES code and a brief description of each molecule. To this set we append add some short alkanes and alkenes for use in the construction of more complex molecules by a fragment-based approach, and caffeine as an example of a small molecule drug. These predictions are made for the materials in the development set and the surfaces used for testing, plus the (001) surfaces of platinum, cerium, chromium oxide (\ce{Cr2O3}), tricalcium silicate, hydroxyapatite, a range of gold (001) surface with $5/25/50/75\%$ of the surface atoms randomly removed to mimic weathering of the surface, and gold (001) surfaces modified with a dense rigid brush (100\% grafting density) of either PEG and PE polymers \cite{charmmguinanomodeller}. These latter two are less physically realistic in that they do not allow for motion of the brushes, but may be a reasonable first approximation to the potential which would be obtained for solid polymer NPs since the gold surface is sufficiently far away to not significant contribute to the potentials. We generate a set of PMFs for all material-chemical pairss both matching the original parameters and a ``canonical'' set of PMFs using the parameters equivalent to Stockholm methodology with ions, SSD type 1, $\Delta_H = 0.002$, $\Delta_P = 0$, $r_0 = 0.2$ nm. Both matched and canonical PMFs are included in the repository and the results for the GAFF parameterised biomolecules required for UnitedAtom \cite{power2019multiscale} are copied to a secondary archive for ease of access. Samples of the matched, canonical and metadynamics PMFs for this subset are shown in Figures S3, S4, S5 and S6 of the ESI with all plots available in the repository \cite{rousepmfparchive}.

\section{Discussion}
The model and methodology proposed here allow for a cost-effective prediction of interactions for multiple classes of materials and chemicals. When trained on a sufficiently diverse set of materials and chemicals, they will be able provide a universal tool for a fast screening of adsorbates in silico. Yet, they are by no means definitive and further optimisation is possible, both for the selection of probes used to define the chemicals and surfaces, and the structure of the network used. In this work, we have attempted to develop a model that is robust enough to demonstrate the overall methodology while still remaining sufficiently flexible to make predictions for a wide range of surfaces and chemicals despite the inhomogeneity of the pool of input PMFs taken from different sources. Crucially, the input is generic enough that new surfaces and adsorbates can be defined without requiring any retraining of the model. The procedure used to parameterise chemicals and materials relies on the existence of a set of co-ordinates and suitable forcefield for the species in question, but provided these exist or can be obtained then the PMFs generated should be valid for a wide range of material surfaces, including high-order Miller indices, amorphous structures, or crystal planes with missing or adsorbed atoms. This is a consequence of the fact that the methodology relies on the construction of the free energy as a function of the distance from the surface and does not directly attempt to make predictions based on the exact structure or component atoms. Thus, as long as the input potentials are physically realistic and not too dissimilar from those in the training set, the output PMF should at least be a reasonable approximation to the one which would be obtained after performing metadynamics simulations with substantially less time investment. The potentials and HGE coefficients produced are themselves useful descriptors of materials for further use in advanced applications.

A key challenge that we tried to answer in building the predictive model is the size and reliability of the dataset used to train it. Here, only a fairly limited number of materials and chemicals have been considered, and the training dataset lacks consistency in terms of the results for nominally the same surface and chemical, e.g. the differing PMFs provided for the \ce{Au} (100) surface which in certain cases produce substantially different results between groups. Moreover,  PMFs produced by the same group for closely related surfaces, e.g. Au (100) and Au (111), exhibit surprising differences despite nearly identical input structures (ESI Figure S2). Since the exact reason for these inconsistencies is not known, we are limited to labelling the dataset by methodology, and it is further not known if these differences reflect genuine differences between the simulated systems or errors in the metadynamics calculations or associated metadata.  Further complications arise due to the inconsistent definitions of the location of the surface and the definition of surface-adsorbate distance. Again, we have attempted to develop the methodology to compensate for this, but a more standardised definition would be beneficial for future work. Another small error has been introduced due to the use of an incorrect structure for the CHARMM parameterisation of GANSCA/GLUPSCA, but this impacts only six training PMFs and is compensated for by the GAFF version for other PMFs and so is unlikely to constitute a large source of error.

Despite these limitations, we generally find a good agreement between the adsorption energies predicted using the model presented here and those found through computationally demanding metadynamics simulations, even for materials not included in the training set. We observe that the model remains generally accurate for these new materials over the range of binding energies in the training set but has difficult extrapolating to even more strongly binding materials. The limited data for testing new chemicals makes it difficult to evaluate whether this limit is responsible for the poor performance of chemicals in the testing set and this remains a topic for further study. The first testing chemical BGALNA is an epimer of the training chemical  BGLCNA, yet exhibits significantly different metadynamics results (ESI Fig S2). The second testing chemical is correctly predicted by some members of the ensemble but not others, suggesting that this may require a larger ensemble or a greater proportion of training data in each ensemble member. For chemicals or surfaces for which no reference is available, we recommend inspection of the PMFs predicted by individual ensembles and the distribution of associated binding energies, under the assumption that if all ensemble members predict the same binding energy this is likely to be more reliable than if there is a significant spread.

For future use, the model may be fine-tuned for a specific material or chemical through use of transfer learning by generating PMFs for a limited number of examples and re-training the model using a very low learning rate for a small number of epochs \cite{pml1Book}.  Based on the typical success of transfer learning it is likely that this would enable the model to make reliable predictions for a novel structures within a much shorter period of time than would be necessary to generate a full set of PMFs. Even without this step, the predictions for materials similar to those in the training set are likely to be quite accurate given the generally good performance for the validation set, especially for perturbations of existing structures, e.g., introducing surface defects into an otherwise pristine structure or modifications of the charge of surface atoms. For organic molecules, ACPYPE and the CHARMM-GUI ligand generator provide well-tested means to generate the input structures and atomic parameters required for essentially arbitrary molecules. Provided that these molecules are similar to those in the training set i.e., organic molecules with low formal charges and molecular masses under 200 amu, it is reasonable to assume that the predictions of the model will be at least approximately correct. At present, the model cannot accurately account for flexibility of input molecules, which may explain some limitations in the reliability of the model. We intend to improve this in future versions of the model by representing molecules as an ensemble of structures rather than the single structure currently used, similar to recent work on the SPICE dataset \cite{eastman2022spice}. This also offers the scope to expand the model to larger chemicals and flexible surfaces such as brushes.

\section{Conclusions}
We have developed a flexible framework for the prediction of PMFs of the interaction between small organic molecules and solid surfaces using a combination of atomistic properties and an artificial neural network. Our methodology represents complex input structures in terms of a universal expansion into basic interaction potentials which can be generated for new molecules and surfaces using their structures and molecular dynamics forcefield parameters. We find a generally good agreement between the target and predicted PMFs in both training and validation sets. Our model enables the rapid analysis of a complex surface in terms of its activity towards small molecules, with applications in catalysis, drug design and computational nanotoxicity.

\section*{Author Contributions}
I. Rouse: Conceptualization, Methodology, Software, Validation, Formal analysis, Investigation, Data Curation, Writing - Original Draft, Writing - Review \& Editing, Visualization. 
V. Lobaskin: Conceptualization, Methodology, Resources, Writing - Review \& Editing, Supervision, Project administration, Funding acquisition.

\section*{Conflicts of interest}
There are no conflicts to declare.

\section*{Acknowledgements}
We thank J. Subbotina, P. Mosaddeghi and A. Lyubartsev for providing details of the metadynamics simulations, and K. Kotsis, A. Colibaba  and the members of QSAR Lab (Gdansk) for calculating nanomaterial descriptors used in development and for useful discussions. We acknowledge funding from EU Horizon 2020 grants No. 814572 (NanoSolveIT) and No. 101008099 (CompSafeNano). 





\bibliography{refs} 
\bibliographystyle{rsc} 

\end{document}


\title{ESI for Machine-learning based prediction of small molecule – surface interaction potentials}
\author{Ian Rouse$^{\ast}$\textit{$^{a}$} \and Vladimir Lobaskin \textit{$^{a }$} }
\maketitle


\renewcommand*\rmdefault{bch}\normalfont\upshape
\rmfamily
\section*{}
\vspace{-1cm}

\renewcommand{\thetable}{S\arabic{table}}
\renewcommand{\thefigure}{S\arabic{figure}}



\section{Additional information}
\begin{figure}[tb]
     \centering
     \includegraphics[width=0.9\textwidth]{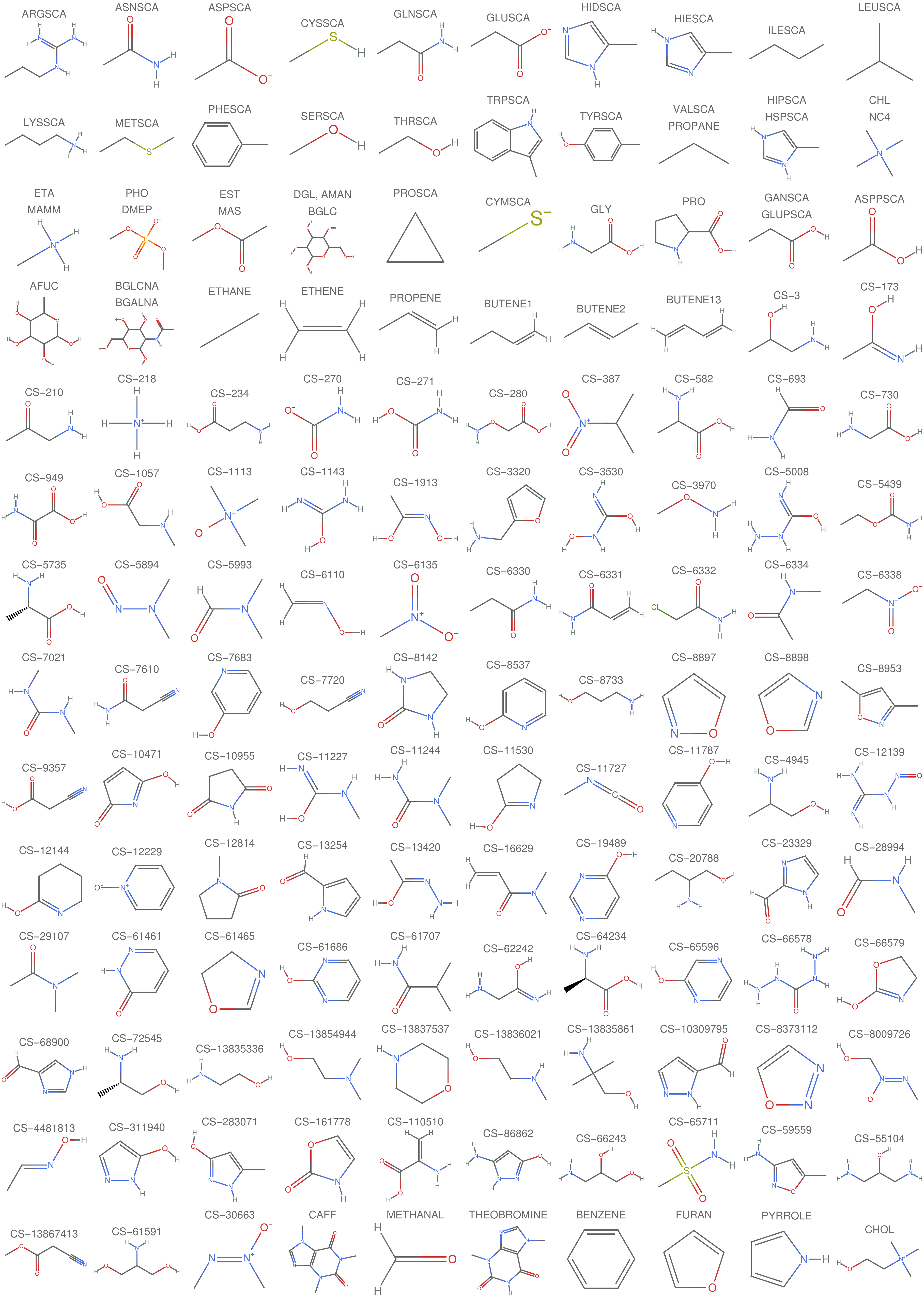}
     \caption{ID tags and structures for the full set of chemicals parameterised for the prediction of PMFs. Molecules sharing a SMILES code are grouped but may possess different structures or force field parameterisations.}
     \label{fig:allchemstructures}
 \end{figure}

\begin{figure}[tb]
     \centering
     \includegraphics[width=0.95\textwidth]{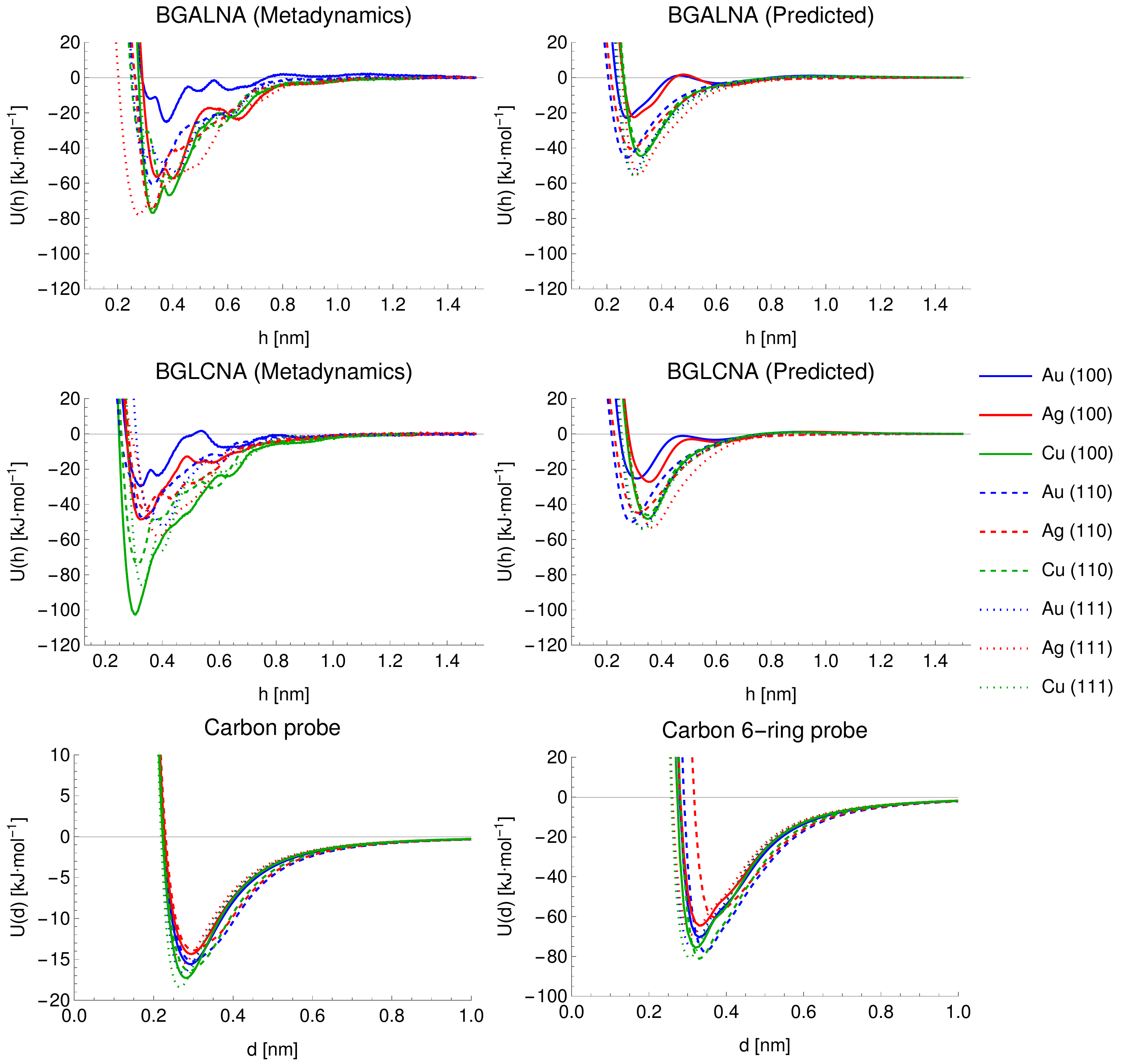}
     \caption{Top: Potentials of mean force for the testing sugar molecule BGALNA on nine FCC metal surfaces obtained via metadyamics simulations (left) and using the approach detailed in this work (right). Prediction parameters are matched to those used for the metadynamics simulations. Middle: As top, but for the sugar molecule BGLCNA used in the training set. Bottom: Input potentials for these FCC metal surfaces showing the single atom carbon probe (left) and the six-membered ring probe (right). In all cases, the gold (blue) and silver (red) surfaces are used in training while copper (green) was reserved for testing.}
     \label{fig:bgalna}
 \end{figure}

In Table \ref{tab:trainingChemicals} we provide SMILES codes and ID tags for all the molecules used in the development of the PMF prediction model, with the additional chemicals generated by selection of highly-cited SMILES codes from ChemSpider and miscellanous small fragments in \ref{tab:extraChemicalsP1} and \ref{tab:extraChemicalsP2}. Structures for all chemicals and their ID tags are shown in Figure \ref{fig:allchemstructures}, with chemicals sharing a SMILES code grouped together into a single figure. These SMILES codes may be mapped to multiple 3D structures or parameterisations if the same SMILES code is used for multiple epimers, enantiomers, etc or if the same molecule is parameterised using both GAFF and CHARMM. We assign a unique ID to each structure based on their equivalent amino acid for side chain analogues, common name or ChemSpider ID and using the suffix "-AC" to denote GAFF parameterisations or "-JS" for CHARMM. In Table \ref{tab:trainingSurfaces} we list details of the surfaces considered in this work, including additional ones parameterised for the generation of PMFs for structures not yet covered by metadynamics. Table \ref{tab:EAdsStats} presents the $R^2$, Pearson correlation, and mean KL divergence for each of the fifteen trained models using the final version of the methodology to identify which models successfully predict PMFs outside of their training set. The computed binding energies for the class of novel chemicals on FCC metal surfaces are provided in Table \ref{tab:EAdsTMNC}. To illustrate the predictive power of the model for the biochemical set of interest for UnitedAtom, we provide example tables of the predicted adsorption energies of the training molecules to amorphous carbon in Table \ref{tab:EAdsCAmorph3} and the Cu (111) surface in Table  \ref{tab:EAdsCu111}. Figure \ref{fig:bgalna} presents a comparison between the predictions for the testing chemical BGALNA on nine metal surfaces, including three copper surfaces not included in the training set and compares these to the closely related molecule BGLCNA and gives examples of the input potentials for these surfaces. In Figures S3 - S6 we show predicted potentials of mean force for a selection of four surfaces, comparing the input PMFs to the predictions generated using parameters matching the particular set of metadynamics used for their input and the canonical form using a set of consistent parameters and surface location determined from the potential probes for all predictions. We show these predictions for the gold (100) and CdSe surfaces to demonstrate strongly and weakly binding planar surfaces and to a hydrated \ce{TiO2} surface and surface-modified CNT to demonstrate more complex planar and cylindrical surfaces respectively. 

  \begin{table}[b]
\addtolength{\tabcolsep}{-1pt}
\begin{tabular}{cccc}
\hline
ID & Source(s) & SMILES &   Description \\
\hline
ALASCA & All & \text{C} &   Alanine side chain (methane) \\
ARGSCA & All & \text{CCCNC(N)=[NH2+]} & Arginine side chain (1-Propylguanidinium) \\
 \text{ASNSCA} & All & \text{CC(N)=O} &   ASN side chain (acetamide) \\
 \text{ASPSCA} & All & \text{CC(=O)[O-]} &   ASP side chain (acetate) \\
 \text{CYSSCA} & All& \text{CS} &   CYS side chain (carbon sulfide) \\
 \text{GLNSCA} & All & \text{CCC(N)=O} &  GLN side chain (propionamide) \\
 \text{GLUSCA} & All & \text{CCC(=O)[O-]} &   GLU side chain (propionate) \\
 \text{HIDSCA} & All & \text{Cc1cnc[nH]1} &  HID side chain  ($\delta$-4-Methylimidazole) \\
 \text{HIESCA} & All & \text{Cc1c[nH]cn1} &  HIE side chain ($\epsilon$-4-Methylimidazole )\\
 \text{ILESCA} & All & \text{CCCC} &  ILE side chain (butane) \\
 \text{LEUSCA} & All & \text{CC(C)C} &   LEU side chain (isobutane)  \\
 \text{LYSSCA} & All & \text{CCCC[NH3+]} & LYS side chain (butylammonium) \\
 \text{METSCA} & All & \text{CCSC} &   MET side chain (ethyl methyl sulfide) \\
 \text{PHESCA} &  All& \text{Cc1ccccc1} &  PHE side chain (toluene) \\
 \text{SERSCA} & All & \text{CO} & SER side chain (methanol) \\
 \text{THRSCA} & All & \text{CCO} &   THR side chain (ethanol) \\
 \text{TRPSCA} & All & \text{Cc1c[nH]c2ccccc12} &TRP side chain (3-methylindole) \\
 \text{TYRSCA} & All & \text{Cc1ccc(O)cc1} &  TYR side chain (p-cresol) \\
 \text{VALSCA} & All & \text{CCC} &VAL side chain (propane) \\
HIPSCA,HSPSCA & All but SU-1 & \text{Cc1c[nH]c[nH+]1} &  Protonated HIS side chain \\
 PROSCA & UCD & \text{C1CC1} &  Cyclopropane  \\
GANSCA, GLUPSCA & All but SU-1 & \text{CCC(=O)O} & Protonated GLU side chain \\
ASPPSCA & UCD & \text{CC(=O)O} &  Protonated ASP side chain \\
CYM & Both & \text{C[S-]} &  Charged cysteine side chain  \\
 GLY& SU-2 & \text{NCC(=O)O} & Glycine \\
 PRO & SU-2 & \text{O=C(O)C1CCCN1} &  Proline \\
ETA, MAMM & All but SU-1  & \text{C[NH3+]} &  \text{Ethanolamine} \\
 DMEP,PHO & All but SU-1  & \text{COP(=O)([O-])OC} &  Phosphate \\
CHL, NC4 & All but SU-1  & \text{C[N+](C)(C)C} &  Tetramethylammonium\\
 EST,MAS  & All but SU-1  & \text{COC(C)=O} &  Ester linkage \\
AFUC & UCD & \text{CC1OC(O)C(O)C(O)C1O} & Alpha fucose \\
BGLCNA & UCD & \text{CC(=O)NC1C(O)OC(CO)C(O)C1O} &  2-acetyl-2-deoxy-beta-d-glucosamine \\
AMAN & UCD & \text{OCC1OC(O)C(O)C(O)C1O} &  Alpha-d-mannose \\
DGL, BGLC & All but SU-1  & \text{OCC1OC(O)C(O)C(O)C1O} &  Beta-d-glucose  \\
\hline
BGALNA & UCD & \text{CC(=O)NC1C(O)OC(CO)C(O)C1O} &  2-acetyl-2-deoxy-beta-d-galactoseamine \\
CHOL & UCD & \text{C[N+](C)(C)CCO} & Choline \\
 \end{tabular}
     \caption{A summary of the chemicals present in the dataset used to build the model for the prediction of PMFs, containing the ID(s) for that molecule used, which PMF computational methods have been used for molecule, the SMILES code, formal charge, and a comment describing the chemical in question and providing the equivalent chemical for the side-chain analogues. In the toolkit, the ID is modified by ``-JS'' or ``-AC'' to distinguish between structures provided by UCD or generated using Charmm GUI and generated by ACPype respectively. }
    \label{tab:trainingChemicals}
 \end{table}

 \begin{table}[tb]
    \centering
    \begin{tabular}{ccc}
\hline
ID & Smiles & Description \\
\hline
 \text{ETHANE} & \text{CC} & \text{Ethane} \\
 \text{PROPANE} & \text{CCC} & \text{Propane} \\
 \text{ETHENE} & \text{C=C} & \text{Ethene} \\
 \text{PROPENE} & \text{C=CC} & \text{Propene} \\
 \text{BUTENE1} & \text{C=CCC} & \text{But-1-ene} \\
 \text{BUTENE2} & \text{CC=CC} & \text{But-2-ene} \\
 \text{BUTENE13} & \text{C=CC=C} & \text{But-1,3-ene} \\
 \text{CS-3} & \text{CC(CN)O} & \text{1-amino-2-propanol} \\
 \text{CS-173} & \text{CC(=N)O} & \text{Acetamide} \\
 \text{CS-210} & \text{CC(=O)CN} & \text{Aminoacetone} \\
 \text{CS-218} & \text{[NH4+]} & \text{Ammonium ion} \\
 \text{CS-234} & \text{C(CN)C(=O)O} & \text{Beta-alanine} \\
 \text{CS-270} & \text{C(=O)(N)[O-]} & \text{Carbamate} \\
 \text{CS-271} & \text{C(=O)(N)O} & \text{Carbamic acid} \\
 \text{CS-280} & \text{C(C(=O)O)ON} & \text{(Aminooxy)acetic acid} \\
 \text{CS-387} & \text{CC(C)[N+](=O)[O-]} & \text{2-nitropropane} \\
 \text{CS-582} & \text{CC(C(=O)O)N} & \text{DL-alanine} \\
 \text{CS-693} & \text{C(=O)N} & \text{Formamide} \\
 \text{CS-730} & \text{C(C(=O)O)N} & \text{Glycine} \\
 \text{CS-949} & \text{C(=O)(C(=O)O)N} & \text{Oxamate} \\
 \text{CS-1057} & \text{CNCC(=O)O} & \text{Sarcosine} \\
 \text{CS-1113} & \text{C[N+](C)(C)[O-]} & \text{Trimethylamine oxide} \\
 \text{CS-1143} & \text{C(=N)(N)O} & \text{Urea} \\
 \text{CS-1913} & \text{C/C(=N/O)/O} & \text{Acetohydroxamic acid} \\
 \text{CS-3320} & \text{c1cc(oc1)CN} & \text{Furfurylamine} \\
 \text{CS-3530} & \text{C(=N)(NO)O} & \text{Hydroxyurea} \\
 \text{CS-3970} & \text{CON} & \text{Methoxyamine} \\
 \text{CS-5008} & \text{C(=N)(NN)O} & \text{Semicarbazide} \\
 \text{CS-5439} & \text{CCOC(=O)N} & \text{Urethane} \\
 \text{CS-5735} & \text{C[C@@H](C(=O)O)N} & \text{L-alanine} \\
 \text{CS-5894} & \text{CN(C)N=O} & \text{Dimethylnitrosamine} \\
 \text{CS-5993} & \text{CN(C)C=O} & \text{N,N-dimethylformamide} \\
 \text{CS-6110} & \text{C=NO} & \text{Formoxime} \\
 \text{CS-6135} & \text{C[N+](=O)[O-]} & \text{Nitromethane} \\
 \text{CS-6330} & \text{CCC(=O)N} & \text{Propionamide} \\
 \text{CS-6331} & \text{C=CC(=O)N} & \text{Acrylamide} \\
 \text{CS-6332} & \text{C(C(=O)N)Cl} & \text{2-chloroacetamide} \\
 \text{CS-6334} & \text{CC(=O)NC} & \text{N-methylacetamide} \\
 \text{CS-6338} & \text{CC[N+](=O)[O-]} & \text{Nitroethane} \\
 \text{CS-7021} & \text{CNC(=O)NC} & \text{N,N'-dimethylurea} \\
 \text{CS-7610} & \text{C(C$\#$N)C(=O)N} & \text{2-cyanoacetamide} \\
 \text{CS-7683} & \text{c1cc(cnc1)O} & \text{3-Pyridinol} \\
 \text{CS-7720} & \text{C(CO)C$\#$N} & \text{Hydracrylonitrile} \\
 \text{CS-8142} & \text{C1CNC(=O)N1} & \text{2-imidazolidinone} \\
 \text{CS-8537} & \text{c1ccnc(c1)O} & \text{2-pyridone} \\
 \text{CS-8733} & \text{C(CN)CO} & \text{3-amino-1-propanol} \\
 \text{CS-8897} & \text{c1cnoc1} & \text{Isoxazole} \\
 \text{CS-8898} & \text{c1cocn1} & \text{Oxazole} \\
 \text{CS-8953} & \text{Cc1cc(on1)C} & \text{3,5-Dimethylisoxazole} \\
 \text{CS-9357} & \text{C(C$\#$N)C(=O)O} & \text{Cyanoacetic acid} \\
 \text{CS-10471} & \text{C1=CC(=O)N=C1O} & \text{Maleimide} \\
 \text{CS-10955} & \text{O=C1CCC(=O)N1} & \text{Succinimide} \\
 \text{CS-11227} & \text{CNC(=N)O} & \text{1-Methylurea} \\
 \text{CS-11244} & \text{CN(C)C(=O)N} & \text{1,1-dimethylurea} \\
 \text{CS-11530} & \text{C1CC(=NC1)O} & \text{2-Pyrrolidone} \\
    \end{tabular}
    \caption{Additional chemicals generated via their SMILES code using ACPYPE and parameterised for use in the PMF Prediction toolkit, part 1}
    \label{tab:extraChemicalsP1}
\end{table}
 
  \begin{table}[tb]
    \centering
    \begin{tabular}{ccc}
\hline
ID & Smiles & Description \\
\hline
 \text{CS-11727} & \text{CN=C=O} & \text{Methyl isocyanate} \\
 \text{CS-11787} & \text{c1cnccc1O} & \text{4-pyridone} \\
 \text{CS-4945} & \text{CC(CO)N} & \text{DL-Alaninol} \\
 \text{CS-12139} & \text{C(=N)(N)NN=O} & \text{1-Nitrosoguanidine} \\
 \text{CS-12144} & \text{C1CCN=C(C1)O} & \text{2-Piperidone} \\
 \text{CS-12229} & \text{c1cc[n+](cc1)[O-]} & \text{pyridine oxide} \\
 \text{CS-12814} & \text{CN1CCCC1=O} & \text{1-methyl-2-pyrrolidinone} \\
 \text{CS-13254} & \text{c1cc([nH]c1)C=O} & \text{2-Formyl-1H-pyrrole} \\
 \text{CS-13420} & \text{C/C(=N/N)/O} & \text{Acethydrazide} \\
 \text{CS-16629} & \text{CN(C)C(=O)C=C} & \text{N,N-dimethylacrylamide} \\
 \text{CS-19489} & \text{c1cncnc1O} & \text{1H-Pyrimidin-4-one} \\
 \text{CS-20788} & \text{CCC(CO)N} & \text{(+/-)-2-amino-1-butanol} \\
 \text{CS-23329} & \text{c1cnc([nH]1)C=O} & \text{Imidazole-2-carbaldehyde} \\
 \text{CS-28994} & \text{CNC=O} & \text{N-methylformamide} \\
 \text{CS-29107} & \text{CC(=O)N(C)C} & \text{N,N-dimethylacetamide} \\
 \text{CS-61461} & \text{c1cc(=O)[nH]nc1} & \text{Pyridazinone} \\
 \text{CS-61465} & \text{C1COC=N1} & \text{Oxazoline} \\
 \text{CS-61686} & \text{c1cnc(nc1)O} & \text{Pyrimidone} \\
 \text{CS-61707} & \text{CC(C)C(=O)N} & \text{Isobutyramide} \\
 \text{CS-62242} & \text{C(C(=N)O)N} & \text{Glycinamide} \\
 \text{CS-64234} & \text{C[C@H](C(=O)O)N} & \text{D-(-)-Alanine} \\
 \text{CS-65596} & \text{c1cnc(cn1)O} & \text{Pyrazinol} \\
 \text{CS-66578} & \text{C(=O)(NN)NN} & \text{Carbohydrazide} \\
 \text{CS-66579} & \text{C1COC(=N1)O} & \text{Oxazolidinone} \\
 \text{CS-68900} & \text{c1c(nc[nH]1)C=O} & \text{4-Imidazolecarboxaldehyde} \\
 \text{CS-72545} & \text{C[C@@H](CO)N} & \text{(S)-(+)-2-amino-1-propanol} \\
 \text{CS-13835336} & \text{C(CO)N} & \text{2-aminoethanol} \\
 \text{CS-13854944} & \text{CN(C)CCO} & \text{Dimethylethanolamine} \\
 \text{CS-13837537} & \text{C1COCCN1} & \text{Morpholine} \\
 \text{CS-13836021} & \text{CNCCO} & \text{N-methylethanolamine} \\
 \text{CS-13835861} & \text{CC(C)(CO)N} & \text{2-amino-2-methyl-1-propanol} \\
 \text{CS-10309795} & \text{c1cn[nH]c1C=O} & \text{1H-pyrazole-3-carbaldehyde} \\
 \text{CS-8373112} & \text{C1=CON=N1} & \text{Oxadiazole} \\
 \text{CS-8009726} & \text{[O-]$\backslash \backslash $[N+](=N/C)CO} & \text{Hydroxymethyl-methylimino-oxido-ammonium} \\
 \text{CS-4481813} & \text{C/C=N/O} & \text{Acetaldoxime} \\
 \text{CS-311940} & \text{c1cn[nH]c1O} & \text{Pyrazol-5-ol} \\
 \text{CS-283071} & \text{Cc1cc(n[nH]1)O} & \text{3-Methyl-5-pyrazolone} \\
 \text{CS-161778} & \text{c1coc(=O)[nH]1} & \text{Oxazolone} \\
 \text{CS-110510} & \text{C=C(C(=O)O)N} & \text{Dehydroalanine} \\
 \text{CS-86862} & \text{c1c([nH]nc1O)N} & \text{3-Amino-5-pyrazolone} \\
 \text{CS-66243} & \text{C(C(CO)O)N} & \text{1-aminoglycerol} \\
 \text{CS-65711} & \text{CS(=O)(=O)N} & \text{Methanesulfonamide} \\
 \text{CS-59559} & \text{Cc1cc(no1)N} & \text{3-amino-5-methyl-isoxazole} \\
 \text{CS-55104} & \text{C(C(CN)O)N} & \text{1,3-diamino-2-propanol} \\
 \text{CS-13867413} & \text{COC(=O)CC$\#$N} & \text{Methyl cyanoacetate} \\
 \text{CS-61591} & \text{C(C(CO)N)O} & \text{Serinol} \\
 \text{CS-30663} & \text{C/N=[N+]($\backslash \backslash $C)/[O-]} & \text{Azoxymethane} \\
 \text{CAFF} & \text{Cn1cnc2n(C)c(=O)n(C)c(=O)c12} & \text{Caffeine} \\
 \text{Water} & \text{O} & \text{Water, TIP3, H charge only} \\
 \text{WaterUCD} & \text{O} & \text{Water, TIP3, H charge and LJ} \\
 \text{METHANAL} & \text{C=O} & \text{Formaldehyde} \\
 \text{THEOBROMINE} & \text{Cn1cnc2c1c(=O)[nH]c(=O)n2C} & \text{Theobromine} \\
 \text{BENZENE} & \text{c1ccccc1} & \text{Benzene} \\
 \text{FURAN} & \text{c1ccoc1} & \text{Furan} \\
 \text{PYRROLE} & \text{[nH]1cccc1} & \text{Pyrrole} \\
    \end{tabular}
    \caption{Additional chemicals generated via their SMILES code using ACPYPE and parameterised for use in the PMF Prediction toolkit, part 2}
    \label{tab:extraChemicalsP2}
\end{table}

 \begin{table}[tb]
    \centering
    \begin{tabular}{cccccc}
\hline
Surface ID & Shape & Source &  SSD Type ($s$)& Resolution [nm]& Notes \\
\hline
 \text{AuFCC100} & Planar & 0 & 0 &0.0032 & FCC Gold (100)\\
 \text{CdSeWurtzite2-10} & Planar & 0 & 0 &0.0045 & \ce{CdSe} Wurtzite (2-10) \\
  \text{TiO2-ana-100} & Planar & 1 & 1 &0.002 & Hydrated anatase (100) \\
  \text{TiO2-ana-101} & Planar & 1 & 1 &0.002 & Hydrated anatase (101) \\
 \text{TiO2-rut-110} & Planar & 1 & 1 &0.002 & Hydrated rutile (110) \\
 \text{TiO2-rut-100} & Planar & 1 & 1 &0.002 & Hydrated rutile (110) \\  
 \text{Fe2O3-001O} & Planar & 1 & 1 &0.002 & Hydrated iron oxide (001) \\
 \text{SiO2-Quartz} & Planar & 1 & 1 &0.002 & Quartz\\
 \text{SiO2-Amorphous} & Planar & 1 & 1&0.002 & Amorphous silica  \\ 
 \text{CNT15} & Cylindrical & 1 & 0&0.002 & Carbon nanotube, 1.5 nm diameter \\
 \text{CNT15-COOH-30} & Cylindrical & 1 & 0&0.002 & Carbon nanotube, 1.5 nm diameter, $30\%$wt \ce{COOH} modified  \\
 \text{CNT15-COOH-3} & Cylindrical & 1 & 0&0.002 & Carbon nanotube, 1.5 nm diameter, $3\%$wt \ce{COOH} modified\\
 \text{CNT15-COO--10} & Cylindrical & 1 & 0&0.002& Carbon nanotube, 1.5 nm diameter, $10\%$wt \ce{COO-} modified \\
 \text{CNT15-COO--3} & Cylindrical & 1 & 0&0.002 & Carbon nanotube, 1.5 nm diameter, $3\%$wt \ce{COO-} modified\\
 \text{CNT15-NH2-14} & Cylindrical & 1 & 0&0.002& Carbon nanotube, 1.5 nm diameter, $14\%$wt \ce{NH2} modified \\
 \text{CNT15-NH2-2} & Cylindrical & 1 & 0 &0.002& Carbon nanotube, 1.5 nm diameter, $2\%$wt \ce{NH2} modified\\
 \text{CNT15-NH3+-4} & Cylindrical & 1 & 0&0.002& Carbon nanotube, 1.5 nm diameter, $4\%$wt \ce{NH3+} modified \\
 \text{CNT15-NH3+-2} & Cylindrical & 1 & 0 &0.002& Carbon nanotube, 1.5 nm diameter, $2\%$wt \ce{NH3+} modified\\
 \text{CNT15-OH-14} & Cylindrical & 1 & 0&0.002& Carbon nanotube, 1.5 nm diameter, $14\%$wt \ce{OH} modified \\
 \text{CNT15-OH-4} & Cylindrical &1 & 0 &0.002& Carbon nanotube, 1.5 nm diameter, $4\%$wt \ce{OH} modified\\
 \text{graphene} & Planar & 1 & 3 &0.002 & Single-layer graphene\\
 \text{bi-graphene} & Planar & 1 & 3&0.002 & Double-layer graphene \\
 \text{tri-graphene} & Planar & 1 & 3&0.002 & Triple-layer graphene\\
 \text{grapheneoxide} & Planar & 1& 3 &0.002 & Graphene oxide, $30\%$ oxidation \\
 \text{redgrapheneoxide} & Planar & 1 & 3&0.002 & Reduced graphene oxide, $10\%$ oxidation\\
 \text{C-amorph-2} & Planar & 1 & 1&0.002 & Amorphous carbon \\
 \text{AuFCC100UCD} & Planar & 3 & 2 &0.00051 & FCC Gold (100) \\
 \text{AuFCC110UCD} & Planar & 2 & 2&0.00034& FCC Gold (110) \\
 \text{AuFCC111UCD} & Planar & 2 & 2 &0.00029& FCC Gold (111)\\
  \text{Ag100} & Planar & 3 & 2 &0.0018& FCC Silver (100)\\
  \text{Ag110} & Planar & 2 & 2 &0.0018& FCC Silver (101)\\
 \text{Ag111} & Planar & 2 & 2&0.0018 & FCC Silver (111) \\
\hline
Fe001&Planar&3&2&0.0018& FCC Iron (001/100)\\
Fe110&Planar&2&2&0.0019& FCC Iron (110)\\
Fe111&Planar&2&2&0.0019& FCC Iron (111)\\
Cu001&Planar&3&2&0.00030& FCC Copper (001/100)\\
Cu110&Planar&2&2&0.00021& FCC Copper (110)\\
Cu111&Planar&2&2&0.00025& FCC Copper (111)\\
C-amorph-3&Planar&1&1&0.002& Amorphous carbon\\
\hline
AuFCC100-Ablate0&Planar&1&0&0.002&FCC Gold (100), $0\%$ ablation\\
AuFCC100-Ablate5&Planar&1&0&0.002&FCC Gold (100), $5\%$ ablation\\
AuFCC100-Ablate25&Planar&1&0&0.002&FCC Gold (100), $25\%$ ablation\\
AuFCC100-Ablate50&Planar&1&0&0.002&FCC Gold (100), $50\%$ ablation\\
AuFCC100-Ablate75&Planar&1&0&0.002&FCC Gold (100), $75\%$ ablation\\
AuFCC100-Ablate100&Planar&1&0&0.002&FCC Gold (100), $100\%$ ablation\\
GoldBrush&Planar&1&0&0.002 & FCC Gold (100), PE brush, $50\%$ coverage\\
CaO001&Planar&1&0&0.002& FCC Calcium oxide (001)\\
Pt001&Planar&1&0&0.002& FCC Platinum(001/100)\\
TricalciumSilicate001&Planar&1&0&0.002& Tricalcium silicate (001)\\
Au-001-PE&Planar&1&0&0.002& Au (100), PE brush, $100\%$ coverage\\
Au-001-PEG&Planar&1&0&0.002 & Au (100), PEG brush, $100\%$ coverage \\
Al2O3-001&Planar&1&0&0.002 & Aluminium oxide (001)\\
Cr2O3-001&Planar&1&0&0.002 & Chromium oxide (001)\\
Ce-001&Planar&1&0&0.002& FCC Cerium (001)\\
Hydroxyapatite-001&Planar&1&0&0.002&Hydroxyapatite (001)\\

    \end{tabular}
    \caption{A summary of the surfaces considered in this work, listing the internal ID, shape, source (0: SU (no ions), 1: SU (ions), 2: UCD-1 (110 and 111 surfaces), 3: UCD-2 (100 surfaces) ) and the zero-indexed categorical variable defining the convention for the SSD (0: nominal surface - COM distance, 1: minimum surface atom - COM distance, 2: slab-width-adjusted COM - COM distance, 3: surface atom COM - COM distance), with e.g. $s=0$ corresponding to SSD class 1 in the main text The first set indicates development surfaces, the second testing surfaces, and the final are additional surfaces without known PMFs. }
    \label{tab:trainingSurfaces}
\end{table}

\begin{table}[tb]
    \centering
\begin{tabular}{ccccc}
\hline
Model ID & Class & $E_{ads}$ Correlation & $E_{ads}$ $R^2$ & $\langle KL \rangle $  \\ 
\hline
Cluster-A-1&TMTC&1.0&0.98&0.17\\
Cluster-A-1&TMVC&0.9&0.78&0.37\\
Cluster-A-1&VMTC&0.94&0.86&0.57\\
Cluster-A-1&VMVC&0.83&0.62&0.69\\
Cluster-A-2&TMTC&1.0&0.99&0.16\\
Cluster-A-2&TMVC&0.89&0.79&0.33\\
Cluster-A-2&VMTC&0.79&0.44&0.71\\
Cluster-A-2&VMVC&0.83&0.67&0.58\\
Cluster-A-3&TMTC&1.0&0.99&0.16\\
Cluster-A-3&TMVC&0.73&0.5&0.36\\
Cluster-A-3&VMTC&0.7&0.36&0.37\\
Cluster-A-3&VMVC&0.67&0.01&0.33\\
Cluster-A-4&TMTC&1.0&0.98&0.16\\
Cluster-A-4&TMVC&0.82&0.66&0.37\\
Cluster-A-4&VMTC&0.88&0.53&0.55\\
Cluster-A-4&VMVC&0.65&0.34&0.87\\
Cluster-A-5&TMTC&1.0&0.98&0.15\\
Cluster-A-5&TMVC&0.9&0.72&0.43\\
Cluster-A-5&VMTC&0.93&0.83&0.38\\
Cluster-A-5&VMVC&0.88&0.76&0.36\\
Simple-B-1&T&1.0&0.99&0.14\\
Simple-B-1&V&0.91&0.82&0.39\\
Simple-B-2&T&1.0&0.99&0.15\\
Simple-B-2&V&0.91&0.83&0.32\\
Simple-B-3&T&1.0&0.99&0.14\\
Simple-B-3&V&0.9&0.82&0.41\\
Simple-B-4&T&1.0&0.98&0.14\\
Simple-B-4&V&0.89&0.78&0.4\\
Simple-B-5&T&1.0&0.99&0.15\\
Simple-B-5&V&0.91&0.83&0.39\\
Simple-B-6&T&1.0&0.99&0.13\\
Simple-B-6&V&0.94&0.87&0.33\\
Simple-B-7&T&1.0&0.99&0.14\\
Simple-B-7&V&0.93&0.85&0.33\\
Simple-B-8&T&1.0&0.99&0.14\\
Simple-B-8&V&0.87&0.74&0.35\\
Simple-B-9&T&1.0&0.99&0.14\\
Simple-B-9&V&0.92&0.84&0.31\\
Simple-B-10&T&1.0&0.99&0.2\\
Simple-B-10&V&0.93&0.86&0.43\\
\end{tabular}
    \caption{ Summary statistics for the accuracy of binding energies (correlation and $R^2$) and mean Kullback-Leibler divergences for PMFs extracted from the predicted potentials of mean force compared to the input potentials for the training and validation groups for the model versions summarised by a split type (clustering then random or simple random), the use of all data (A) or bootstrap data (B) and an ID number. For simple random splitting, the class column indicates results for PMFs included in the training set for that particular bootstrap dataset and training-validation split, with the validation set containing all out-of-bag PMFs. For cluster splitting, TMTC indicates the training material, training chemical set and so on, with all PMFs featured in one of these four classes.}
    \label{tab:EAdsStats}
\end{table}

\begin{table}[tb]
    \centering
\begin{tabular}{ccccc}
\hline
 \text{Material} & \text{Chemical} & \text{E(MD)} $[\text{kJ}\cdot\text{mol}^{-1}]$  &  E(Simple) $[\text{kJ}\cdot\text{mol}^{-1}]$ &  E(Cluster) $[\text{kJ}\cdot\text{mol}^{-1}]$\\
 \hline
Ag100 & BGALNA-JS &  $-49.$ & $-13.7\pm 2.7$ & $-13.5\pm 2.1$ \\
Ag100 & CHOL-JS &  $-10.6$ & $-7.9\pm 2.9$ & $-3.3\pm 0.8$ \\
Ag110 & BGALNA-JS &  $-64.4$ & $-32.\pm 4.$ & $-31.2\pm 2.9$ \\
Ag110 & CHOL-JS &  $-21.2$ & $-29.\pm 6.$ & $-27.\pm 5.$ \\
Ag111 & BGALNA-JS &  $-68.8$ & $-46.\pm 6.$ & $-43.\pm 5.$ \\
Ag111 & CHOL-JS &  $-15.8$ & $-34.\pm 11.$ & $-38.\pm 14.$ \\
AuFCC100UCD & BGALNA-JS &  $-15.5$ & $-14.\pm 4.$ & $-12.0\pm 2.0$ \\
AuFCC100UCD & CHOL-JS &  $-7.8$ & $-7.\pm 4.$ & $-5.3\pm 0.7$ \\
AuFCC110UCD & BGALNA-JS &  $-51.$ & $-36.\pm 4.$ & $-33.\pm 6.$ \\
AuFCC110UCD & CHOL-JS &  $-20.2$ & $-30.\pm 6.$ & $-32.\pm 13.$ \\
AuFCC111UCD & BGALNA-JS &  $-45.1$ & $-45.\pm 9.$ & $-45.6\pm 2.3$ \\
AuFCC111UCD & CHOL-JS &  $-18.5$ & $-32.\pm 8.$ & $-33.\pm 12.$ \\
Cu001 & BGALNA-JS &  $-67.2$ & $-36.\pm 6.$ & $-31.\pm 4.$ \\
Cu110 & BGALNA-JS &  $-51.3$ & $-33.\pm 9.$ & $-27.5\pm 3.2$ \\
Cu111 & BGALNA-JS &  $-65.3$ & $-46.\pm 8.$ & $-47.6\pm 2.2$ \\

\end{tabular}
    \caption{Adsorption energies of two molecules not used in the training set to FCC gold, silver and copper for (100), (110) and (111) surfaces, comparing the values found from metadynamics simulations to those predicted by the two ensemble models. The provided errors are obtained from the standard deviation of the adsorption energies predicted by each individual ensemble member. }
    \label{tab:EAdsTMNC}
\end{table}

\begin{table}[tb]
    \centering
\begin{tabular}{ccccc}
\hline
 \text{Material} & \text{Chemical} & \text{E(MD)} $[\text{kJ}\cdot\text{mol}^{-1}]$  &  E(Simple) $[\text{kJ}\cdot\text{mol}^{-1}]$ &  E(Cluster) $[\text{kJ}\cdot\text{mol}^{-1}]$ \\
 \hline
 C-amorph-3 & ALASCA-AC &  $-2.2$ & $-0.3\pm 0.5$ & $0.60\pm 0.19$ \\
 C-amorph-3 & ARGSCA-AC &  $-16.5$ & $-16.9\pm 1.1$ & $-11.7\pm 2.3$ \\
 C-amorph-3 & ASNSCA-AC &  $-8.5$ & $-6.9\pm 1.5$ & $-5.2\pm 2.0$ \\
 C-amorph-3 & ASPSCA-AC &  $-4.1$ & $-2.3\pm 1.2$ & $-1.8\pm 1.0$ \\
 C-amorph-3 & CHL-AC &  $-3.1$ & $-2.1\pm 0.9$ & $-1.47\pm 0.33$ \\
 C-amorph-3 & CYMSCA-AC &  $-2.2$ & $-1.5\pm 1.2$ & $-0.9\pm 0.8$ \\
 C-amorph-3 & CYSSCA-AC &  $-5.1$ & $-2.0\pm 0.8$ & $-0.61\pm 0.33$ \\
 C-amorph-3 & DGL-AC &  $-19.$ & $-17.1\pm 2.6$ & $-14.\pm 4.$ \\
 C-amorph-3 & EST-AC &  $-13.5$ & $-12.4\pm 1.4$ & $-9.2\pm 1.6$ \\
 C-amorph-3 & ETA-AC &  $-0.7$ & $0.0\pm 0.5$ & $0.35\pm 0.33$ \\
 C-amorph-3 & GANSCA-AC &  $-12.8$ & $-11.9\pm 1.6$ & $-8.7\pm 1.6$ \\
 C-amorph-3 & GLNSCA-AC &  $-11.3$ & $-9.7\pm 1.7$ & $-7.2\pm 1.3$ \\
 C-amorph-3 & GLUSCA-AC &  $-6.5$ & $-7.1\pm 3.0$ & $-5.6\pm 2.1$ \\
 C-amorph-3 & GLY-AC &  $-9.8$ & $-8.4\pm 1.9$ & $-7.4\pm 2.5$ \\
 C-amorph-3 & HIDSCA-AC &  $-14.8$ & $-14.9\pm 1.2$ & $-11.5\pm 2.5$ \\
 C-amorph-3 & HIESCA-AC &  $-14.9$ & $-15.9\pm 0.8$ & $-12.9\pm 1.9$ \\
 C-amorph-3 & HIPSCA-AC &  $-12.7$ & $-15.3\pm 1.7$ & $-11.\pm 4.$ \\
 C-amorph-3 & ILESCA-AC &  $-11.9$ & $-8.0\pm 1.9$ & $-5.9\pm 1.6$ \\
 C-amorph-3 & LEUSCA-AC &  $-11.8$ & $-7.3\pm 2.7$ & $-5.6\pm 0.9$ \\
 C-amorph-3 & LYSSCA-AC &  $-9.$ & $-5.3\pm 1.7$ & $-3.8\pm 1.4$ \\
 C-amorph-3 & METSCA-AC &  $-12.4$ & $-9.5\pm 2.1$ & $-7.5\pm 2.4$ \\
 C-amorph-3 & PHESCA-AC &  $-16.5$ & $-15.7\pm 2.4$ & $-10.7\pm 0.5$ \\
 C-amorph-3 & PHO-AC &  $-8.$ & $-6.5\pm 2.7$ & $-6.8\pm 2.3$ \\
 C-amorph-3 & PRO-AC &  $-18.4$ & $-15.6\pm 2.4$ & $-11.3\pm 2.7$ \\
 C-amorph-3 & SERSCA-AC &  $-3.9$ & $-1.2\pm 0.7$ & $0.0\pm 0.4$ \\
 C-amorph-3 & THRSCA-AC &  $-6.9$ & $-3.1\pm 0.9$ & $-2.6\pm 1.0$ \\
 C-amorph-3 & TRPSCA-AC &  $-21.1$ & $-25.3\pm 1.9$ & $-19.0\pm 2.5$ \\
 C-amorph-3 & TYRSCA-AC &  $-20.2$ & $-22.6\pm 2.8$ & $-15.5\pm 3.0$ \\
 C-amorph-3 & VALSCA-AC &  $-8.8$ & $-5.1\pm 1.6$ & $-3.0\pm 1.3$ \\
 
\end{tabular}
    \caption{Adsorption energies of molecules used for the development of the model to an amorphous carbon surface not included in the training set, comparing the values found from metadynamics simulations to those predicted by the two ensemble models.  The provided errors are obtained from the standard deviation of the adsorption energies predicted by each individual ensemble member.}
    \label{tab:EAdsCAmorph3}
\end{table}

\begin{table}[tb]
    \centering
\begin{tabular}{ccccc}
\hline
 \text{Material} & \text{Chemical} & \text{E(MD)} $[\text{kJ}\cdot\text{mol}^{-1}]$  &  E(Simple) $[\text{kJ}\cdot\text{mol}^{-1}]$ &  E(Cluster) $[\text{kJ}\cdot\text{mol}^{-1}]$ \\
 \hline
Cu111 & AFUC-JS &  $-43.1$ & $-37.\pm 10.$ & $-39.\pm 5.$ \\
Cu111 & ALASCA-JS &  $1.7$ & $-2.3\pm 2.6$ & $0.4\pm 0.6$ \\
Cu111 & AMAN-JS &  $-55.7$ & $-41.\pm 10.$ & $-55.\pm 5.$ \\
Cu111 & ARGSCA-JS &  $-72.$ & $-52.\pm 12.$ & $-53.\pm 18.$ \\
Cu111 & ASNSCA-JS &  $-20.1$ & $-19.0\pm 2.0$ & $-18.4\pm 1.2$ \\
Cu111 & ASPPSCA-JS &  $-23.6$ & $-18.4\pm 1.7$ & $-17.6\pm 1.4$ \\
Cu111 & ASPSCA-JS &  $-12.2$ & $-10.4\pm 1.7$ & $-9.5\pm 0.4$ \\
Cu111 & BGALNA-JS &  $-65.3$ & $-46.\pm 8.$ & $-47.6\pm 2.2$ \\
Cu111 & BGLC-JS &  $-55.9$ & $-48.\pm 9.$ & $-56.3\pm 3.4$ \\
Cu111 & BGLCNA-JS &  $-76.$ & $-45.\pm 6.$ & $-44.\pm 5.$ \\
Cu111 & CYMSCA-JS &  $-22.3$ & $-18.0\pm 1.6$ & $-17.2\pm 0.9$ \\
Cu111 & CYSSCA-JS &  $-14.8$ & $-13.8\pm 1.3$ & $-12.0\pm 0.9$ \\
Cu111 & DMEP-JS &  $-57.5$ & $-39.\pm 6.$ & $-45.4\pm 2.9$ \\
Cu111 & GLNSCA-JS &  $-25.7$ & $-21.5\pm 1.9$ & $-21.6\pm 1.1$ \\
Cu111 & GLUPSCA-JS &  $-29.9$ & $-20.8\pm 2.2$ & $-20.2\pm 2.8$ \\
Cu111 & GLUSCA-JS &  $-14.8$ & $-14.\pm 4.$ & $-11.1\pm 3.0$ \\
Cu111 & HIDSCA-JS &  $-31.8$ & $-27.\pm 4.$ & $-26.2\pm 1.3$ \\
Cu111 & HIESCA-JS &  $-33.6$ & $-26.3\pm 2.8$ & $-24.3\pm 1.5$ \\
Cu111 & HIPSCA-JS &  $-42.$ & $-30.\pm 4.$ & $-29.\pm 5.$ \\
Cu111 & ILESCA-JS &  $-10.1$ & $-12.0\pm 2.5$ & $-12.4\pm 3.4$ \\
Cu111 & LEUSCA-JS &  $-4.$ & $-9.8\pm 3.1$ & $-8.1\pm 1.4$ \\
Cu111 & LYSSCA-JS &  $-12.3$ & $-18.\pm 7.$ & $-15.1\pm 3.1$ \\
Cu111 & MAMM-JS &  $-3.8$ & $-2.8\pm 1.3$ & $-2.2\pm 2.1$ \\
Cu111 & MAS-JS &  $-23.6$ & $-20.3\pm 2.5$ & $-18.9\pm 2.8$ \\
Cu111 & METSCA-JS &  $-27.8$ & $-23.9\pm 2.3$ & $-23.0\pm 0.9$ \\
Cu111 & NC4-JS &  $-14.3$ & $-15.0\pm 2.0$ & $-13.0\pm 1.4$ \\
Cu111 & PHESCA-JS &  $-32.7$ & $-28.\pm 4.$ & $-27.8\pm 1.0$ \\
Cu111 & PROSCA-JS &  $-4.$ & $-8.6\pm 1.5$ & $-6.8\pm 1.0$ \\
Cu111 & SERSCA-JS &  $-3.$ & $-5.0\pm 3.1$ & $-3.4\pm 2.4$ \\
Cu111 & THRSCA-JS &  $-5.9$ & $-9.\pm 4.$ & $-8.6\pm 1.7$ \\
Cu111 & TRPSCA-JS &  $-75.$ & $-54.\pm 11.$ & $-60.6\pm 3.1$ \\
Cu111 & TYRSCA-JS &  $-51.1$ & $-35.\pm 8.$ & $-38.\pm 9.$ \\
Cu111 & VALSCA-JS &  $-1.6$ & $-8.3\pm 2.7$ & $-6.2\pm 0.8$ \\
\end{tabular}
    \caption{Adsorption energies of molecules used for the development of the model to a copper (111) surface not included in the training set, comparing the values found from metadynamics simulations to those predicted by the two ensemble models.  The provided errors are obtained from the standard deviation of the adsorption energies predicted by each individual ensemble member.}
    \label{tab:EAdsCu111}
\end{table}

  \begin{figure}[tb]
     \centering
     \includegraphics[width=0.9\textwidth]{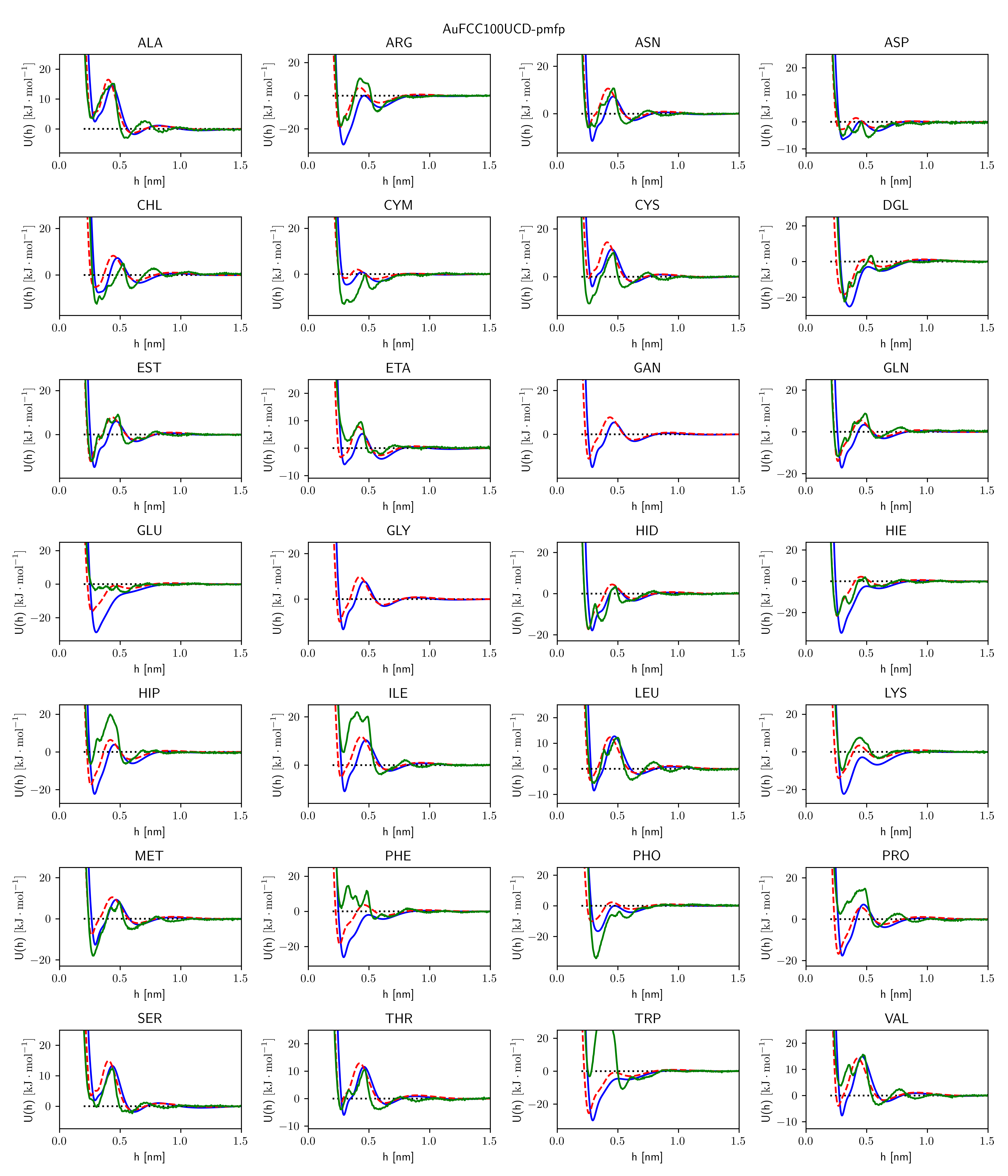}
     \caption{Generated potentials of mean force for the Au FCC (100) surface, showing the canonical form (blue), match to original parameters (red, dashed) and input PMF where available (green).}
     \label{fig:gold}
 \end{figure}

  \begin{figure}[tb]
     \centering
     \includegraphics[width=0.9\textwidth]{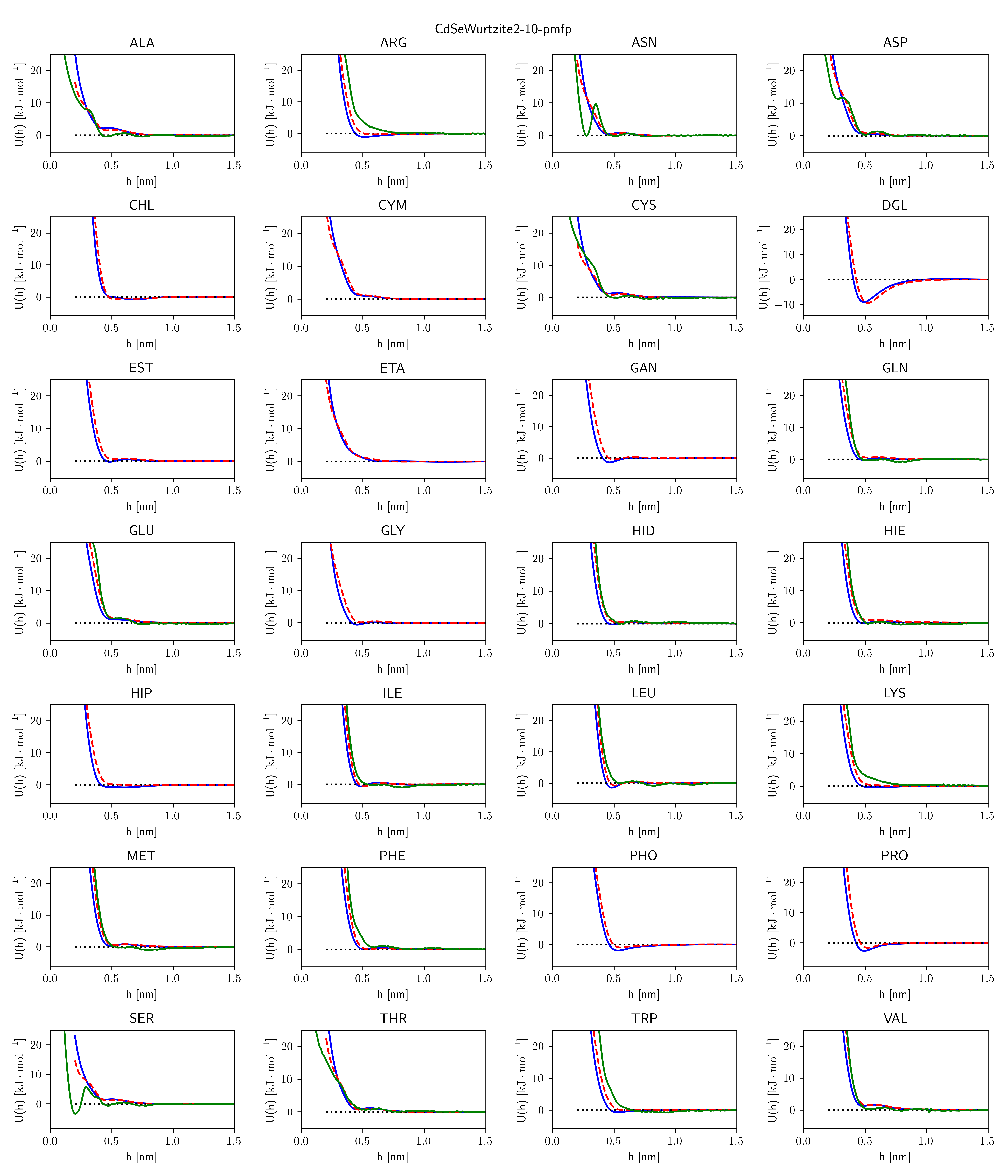}
     \caption{Generated potentials of mean force for the CdSe wurtzite (2-10) surface, showing the canonical form (blue), match to original parameters (red, dashed) and input PMF where available (green).}
     \label{fig:cdse}
 \end{figure}
   \begin{figure}[tb]
     \centering
     \includegraphics[width=0.9\textwidth]{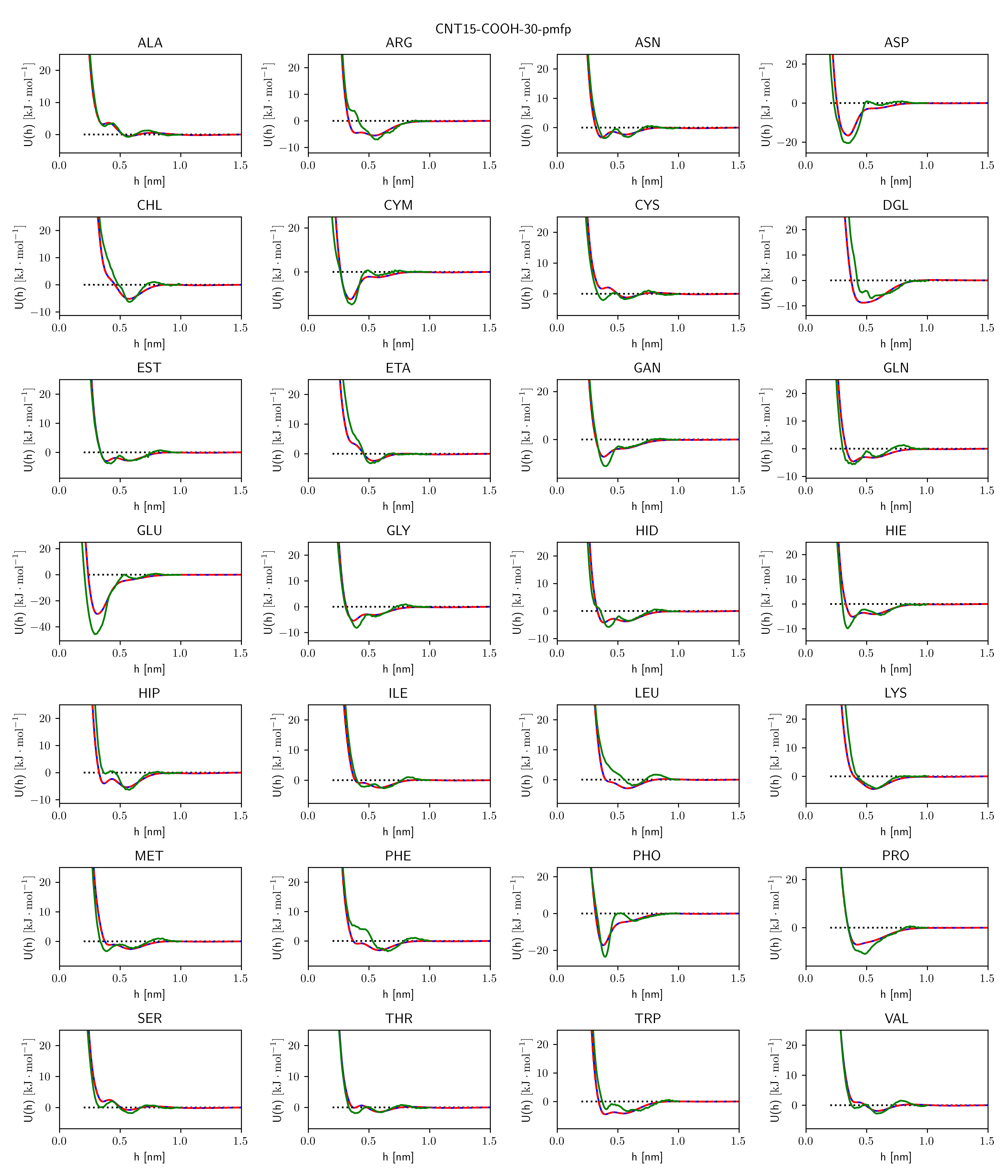}
     \caption{Generated potentials of mean force for the $30\% $ by weight \ce{COOH} modified carbon nanotube surface, showing the canonical form (blue), match to original parameters (red, dashed) and input PMF where available (green).}
     \label{fig:cntcooh30}
 \end{figure}
   \begin{figure}[tb]
     \centering
     \includegraphics[width=0.9\textwidth]{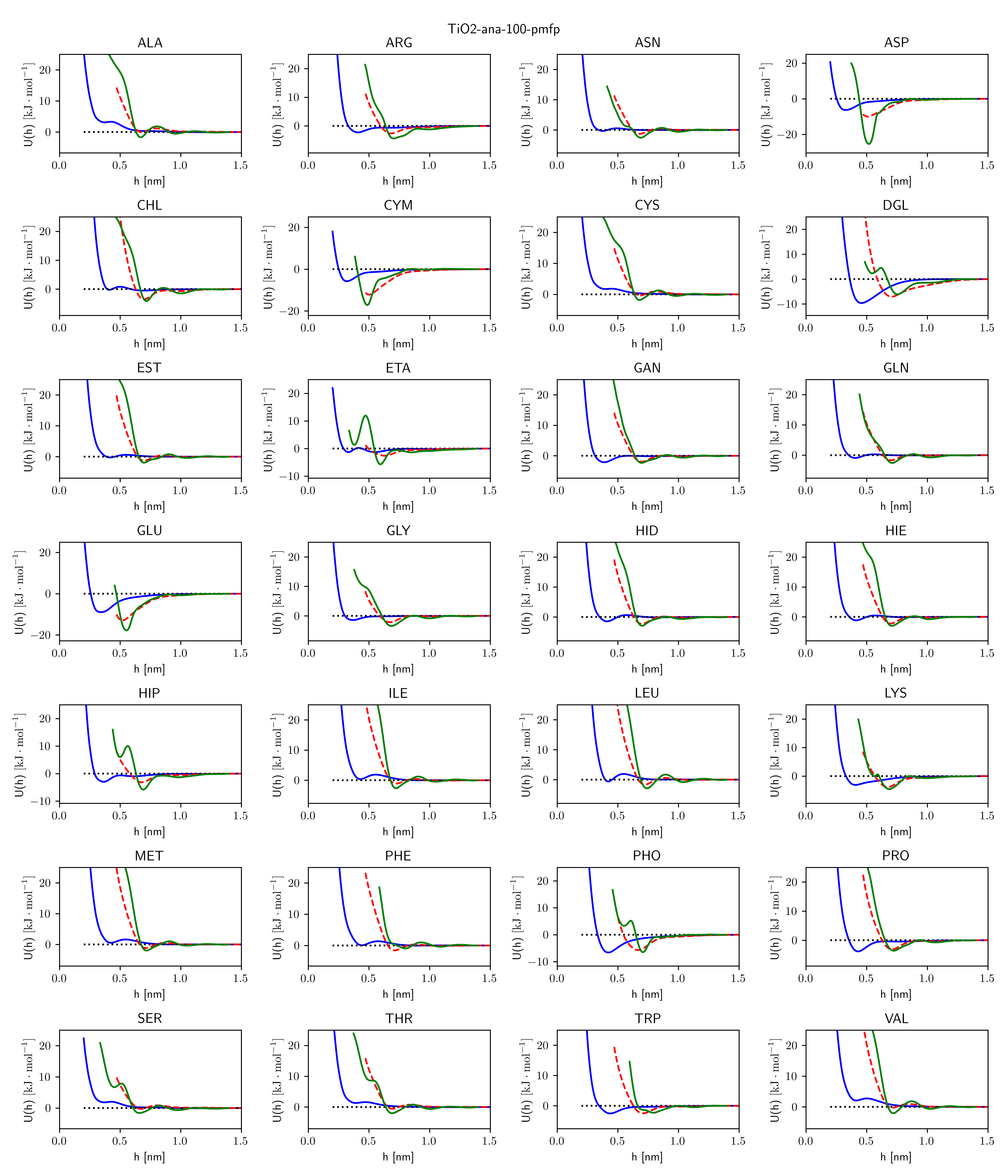}
     \caption{Generated potentials of mean force for the titanium dioxide anatase (100) surface, showing the canonical form (blue), match to original parameters (red, dashed) and input PMF where available (green).}
     \label{fig:tio2ana100}
 \end{figure}
